\documentclass[12pt]{iopart}

\usepackage{graphicx}
\usepackage{dcolumn}
\usepackage{multirow}
\usepackage{placeins}
\usepackage[title]{appendix}

\begin{document}

\title{Magnetism between magnetic adatoms on monolayer NbSe$_2$}

\author{S. Sarkar$^1$, F. Cossu$^{1,2}$, P. Kumari$^{3}$, A.~G. Moghaddam$^{4}$, A. Akbari$^{1,5,6}$, Y.~O. Kvashnin$^7$ and I. {Di Marco}$^{1,7,8}$}

\address{$^1$ Asia Pacific Center for Theoretical Physics, Pohang 37673, Korea}
\address{$^2$ Department of Physics, Kangwon National University – Chuncheon, 24341, Korea}
\address{$^3$ Istituto di Chimica dei Composti Organometallici, Consiglio Nazionale delle Ricerche (ICCOM-CNR), via G. Moruzzi 1, Pisa 56124, Italy}
\address{$^4$ Department of Physics, Institute for Advanced Studies in Basic Sciences (IASBS), Zanjan 45137-66731, Iran}
\address{$^5$ Max Planck Institute for the Chemical Physics of Solids, D-01187 Dresden, Germany}
\address{$^6$ Max Planck POSTECH Center for Complex Phase Materials, and Department of Physics, POSTECH, Pohang, Gyeongbuk 790-784, Korea}
\address{$^7$ Department of Physics and Astronomy, Uppsala University, Box 516, SE-75120 Uppsala, Sweden}
\address{$^8$ Department of Physics, POSTECH, Pohang 37673, Korea}
\ead{igor.dimarco@apctp.org}

\begin{abstract}  
In this work, we report on an {\it{ab-initio}} computational study of the electronic and magnetic properties of transition metal adatoms on a monolayer of NbSe$_2$. We demonstrate that Cr, Mn, Fe and Co prefer all to sit above the Nb atom, where the $d$ states experience a substantial hybridization. The inter-atomic exchange coupling is shown to have an oscillatory nature accompanied by an exponential decay, in accordance with what theory predicts for a damped Ruderman-Kittel-Kasuya-Yosida (RKKY) interaction. Our results indicate that the qualitative features of the magnetic coupling for the four investigated adatoms can be connected to the fine details of their Fermi surface. In particular, the oscillations of the exchange in Fe and Co are found to be related to a single nesting vector, connecting large electrons and hole pockets. Most interestingly, this behavior is found to be unaffected by changes induced on the height of the impurity, which makes the magnetism robust to external perturbations. Considering that NbSe$_2$ is a superconductor down to a single layer, our research might open the path for further research into the interplay between magnetic and superconducting characteristics, which could lead to novel superconductivity engineering. 
\end{abstract}


\submitto{}
\maketitle

\section{Introduction}
The recent years have seen a drastic increase in the research on two-dimensional (2D) magnetism, motivated by a fascinating fundamental physics as well as the perspective of various applications. According to the Mermin-Wagner theorem, long-range magnetic ordering in 2D systems should be suppressed by thermal fluctuations~\cite{mermin66prl17:1133}. Nevertheless, this restriction is inferred from simple models and, in the physical world, it can easily be removed by the presence of a sizeable magnetic anisotropy. Layered van der Waals (vdW) materials have a reduced crystal symmetry and hence are ideal, potential candidates for 2D magnetism~\cite{carteaux95epl29:251}. In fact, several recent studies reported on the discovery of ferromagnetism in vdW materials down to the truly 2D limit~\cite{huang17nature546:270,gong17nature546:265,deng18nature653:94,kim19prl122:207201}. Intrinsic magnetism is naturally the most attractive option in this regard and was confirmed in CrI$_3$, Cr$_2$Ge$_2$Te$_6$ and VSe$_2$, albeit only for the latter the ordering temperature was found to be above room temperature. Unfortunately, structural degradation due to exposure to air has proven to be detrimental for the application of these materials in devices~\cite{sethulakshmi19mtoday27:107}. An alternative strategy to avoid this problem is to work with non-magnetic 2D materials and induce magnetism via the functionalization with defects, doping or adatoms~\cite{novoselov07natmat6:183,xia09natnano4:839,katsnelson12graphenebook,mak14natnano344:1489,slota18nature557:7707,reed21advmatint2100463}. Transition metal dichalcogenides (TMD) occupy a special place among 2D materials, due to their high versatility, tunability, ease of synthesis and manipulation~\cite{manzeli17natrevmat2:1}. Intrinsic as well as extrinsic magnetism has been predicted by theory and confirmed in experiment~\cite{huang21smartmat2:139}. A substantial amount of research on extrinsic magnetism has been focused on insulating systems, as e.g. MoS$_2$ and MoSe$_2$, due to the possibility of realizing localized impurity states with large atomic-like magnetic moments~\cite{huang13jap114:083706,wang14pe63:276,li17natcomm8:1958,wei17prb95:075419,coelho19aem5:1900044,zhang20advsci7:2001174,yun20advsci7:1903076,pham20advmat32:2003607}. Metallic TMDs in 2D are also very interesting, due to the presence of fascinating collective phenomena such as superconductivity and charge density waves (CDWs)~\cite{shi18ccr376:1}. 
A 2D material that exhibits this type of physics is NbSe$_2$, whose structural and electronic properties have been extensively characterized in recent years~\cite{ugeda16natphys12:92,gye19prl122:016403,oh20prl125:036804,soumyanarayanan13pnas110:1623,gao18pnas115:6986,cossu20asiamat12:1,sohn18natmat17:504,hamill21natphys17:949}. A peculiar feature of NbSe$_2$ in the monolayer limit is that it is close to a magnetic instability, which may easily be triggered by external means. For example, long-range order has been shown to arise under a small amount of tensile strain, albeit the precise amount of required strain and the type of magnetic pattern are still debated~\cite{zhou12acsnano6:9727,xu14nanoscale6:12929,zheng18prb97:081101,divilov21jpcm33:295804}. 
Functionalization via doping or adsorbed atoms and molecules has been also explored. 
Zhu {\it{et al.}} have shown in an experiment that superconductivity and ferromagnetism coexist in NbSe$_2$ after adsorption of hydrazine molecules~\cite{zhu16natcomm7:1}. 
Liebhaber {\it{et al.}} have used scanning tunneling microscopy (STM) and electronic structure theory to show that Fe adatoms on NbSe$_2$ lead to the formation of Yu-Shiba-Rusinov bound state with the charge-density order~\cite{liebhaber19nanolet20:339}. Similar analyses for thin films of NbSe$_2$ were also recently published~\cite{kezilebieke18nanolet18:2311,pervin19jms54:11903}. 

In a previous work~\cite{cossu18prb98:195419}, we investigated how the presence of magnetic and non-magnetic adatoms alter the local patterns and energy hierarchy of CDWs in NbSe$_2$. 
In this work, instead, we intend to focus on the very nature of the exchange mechanism driving the formation of the magnetic order in NbSe$_2$ in presence of selected magnetic adatoms, namely Cr, Mn, Fe and Co. By means of {\it{ab-initio}} electronic structure calculations based on density-functional theory (DFT)~\cite{martin_book}, we analyze the magnetic landscape associated to different positions of adatoms on the NbSe$_2$ monolayer, showing the presence of several stable minima at different heights. After obtaining the ground state structures, we extract the inter-atomic exchange interactions between adatoms and show that they have a damped oscillatory nature, which is caused by the Ruderman-Kittel-Kasuya-Yosida (RKKY)-type coupling~\cite{ruderman54pr96:99,pajda01prb64_174402}. 
Most importantly, the qualitative features of the magnetic coupling can be connected to the details of the Fermi surface (FS). For Fe and Co, the presence of a nesting vector connecting large electron and hole pockets is found to essentially determine the oscillatory character of the magnetic interactions. Conversely, the absence of this features for Cr and Mn gives rise to a coupling of a much shorter range. 

The rest of the paper is organized as follows. In Section~\ref{Methodology}, the methodological aspects of our study are presented. Then, Section~\ref{StructureMagnetism} is centered on the analysis of the interplay between geometry and magnetism. The analysis of the long-range character of the exchange coupling is reported in Section~\ref{ExchangeCoupling}. Finally, the conclusions of this work are discussed in Section~\ref{Conclusions}.

\section{Methodology}\label{Methodology}
\subsection{Structural Optimization}
The electronic structure was calculated using a projected augmented wave (PAW) method~\cite{blochl94prb50:17953,kresse99prb59:1758} as implemented in the Vienna ab-initio simulation package 
(VASP)~\cite{kresse93prb47:558,kresse94prb49:14251,kresse96prb54:11169,kresse96cms6:15}. The generalized gradient approximation (GGA)~\cite{perdew96prl77:3865} in the Perdew-Burke-Ernzerhof (PBE) realization~\cite{perdew96prl77:3865,perdew97prl78:1396} was used for the exchange-correlation functional. 
The system to investigate consisted of a $4\times4$ supercell of 1H-NbSe$_2$ monolayer, including a total of 48 atoms, with a single magnetic adatom on top. The magnetic adatoms addressed in our study were  Cr, Mn, Fe, and Co. A vacuum region of 14 \r{A} was considered between two monolayers in the adjacent supercells along the $z$-direction to eliminate any unphysical inter-layer interactions. The in-plane lattice constant was fixed to the experimental value of 3.45 \r{A} corresponding to the NbSe$_2$ unit cell~\cite{kalikhman83inmat19:957}. The internal positions of the atoms were optimized towards the minimum energy configuration until the forces were found to be less than 10$^{-3}$ eV/\r{A}. A gamma centered
Monkhorst-Pack mesh of $19\times19\times1$ {\bf{k}}-points and a plane wave energy cutoff of 800 eV were used in the calculations. 
Due to the localized nature of the $3d$ states in transition metal (TM) adatoms, previous studies on 2D materials have emphasized the need of including an explicit correction due to the local Coulomb interaction~\cite{wehling11prb84:235110,wei17prb95:075419,li17natcomm8:1958,cossu18prb98:195419,reed21advmatint2100463}. This term was then treated in the Hartree-Fock approximation, via the DFT+U method~\cite{anisimov97jpcm9:10581,kotliar06rmp78:865}. In VASP, we employed the rotationally invariant formulation proposed by Liechtenstein {\it{et al.}}~\cite{lichtenstein95prb52:R5467}. The Coulomb interaction parameters $U$ and $J$ were chosen on the basis of previous studies~\cite{wehling11prb84:235110,cossu18prb98:195419} as 4.0 eV and 0.9 eV respectively.

\subsection{Electronic and magnetic properties}

After performing the structural optimization, we investigated the details of the electronic and magnetic properties by means of an all-electron theory, which provides a more accurate solution, albeit at a higher computational cost~\cite{lejaeghere16science351:6280}. To this aim, we employed the full potential linear muffin-tin orbital (FP-LMTO) method as implemented in the Relativistic Spin Polarized Toolkit (RSPt)~\cite{RSPt_book,rspt_website}. 
The radii of the muffin-tin spheres were set to $2.21\div2.34$, $2.15\div2.27$ and $2.09\div2.18$ a.u. for respectively Nb, Se and TM adatoms. The intervals refer to the small variations when going from Co (smallest values) to Mn and Cr (largest values). 
The basis functions were formally divided in two energy sets, one for the valence states and one for the semi-core states. The former included $5s$ $5p$ $4d$ states for Nb, $4s$ $4p$ $4d$ states for Se, and $4s$ $4p$ $3d$ states for adatoms. The latter included $4s$ $4p$ states Nb, $3d$ states for Se, and $3s$ $3p$ states for adatoms. Three kinetic energy tails were used for the valence $sp$ states, corresponding to the values of -0.1, -2.3 and -1.5 Ry. Only the first two tails were used for all the other basis functions.
The DFT+U correction was applied using the spin and orbital rotationally invariant formulation described in Refs.~\cite{grechnev07prb76:035107,granas12cms55:295}, using a local basis that was constructed from the muffin-tin heads. The double counting correction was based on the fully localized limit (FLL)~\cite{anisimov97jpcm9:767,kotliar06rmp78:865}.
The 4-index U-matrix was constructed using the same Coulomb interaction parameters used in VASP and hence, in absence of spin-orbit coupling, the two formalisms should be absolutely equivalent~\cite{shick05epl69:588}. 
The remaining computational settings, including ${\bf{k}}$-point mesh and integration method, were also set to be in accordance with VASP, in order to ensure the consistence of our results across the manuscript. RSPt was also used to calculate the inter-atomic exchange parameters, by mapping the magnetic excitations onto the following Heisenberg Hamiltonian:
\begin{equation}
\hat{H}=-\sum_{i \neq j}J_{ij}\vec{e}_i \cdot \vec{e}_j
\end{equation}
Here $J_{ij}$ is the exchange interaction between the spins at sites $i$ and $j$, while $\vec{e}_i$ and $\vec{e}_j$ are unit vectors along the magnetization direction at the corresponding site.
The $J_{ij}$'s were calculated by employing the generalized magnetic force theorem~\cite{lichtenstein87jmmm67:65,katsnelson00prb61:8906}, whose implementation in RSPt is described in Ref.~\cite{kvashnin15prb91_125133}. The local basis for the calculation of the $J_{ij}$'s was chosen to be equivalent to the one used in DFT+U, which was described above. 
Finally, the convergence of the inter-atomic exchange parameters up to the precision used in our data analysis required the use of a very fine sampling of the Brillouin Zone, extending up to a mesh consisting of $70 \times 70 \times 1$ {\bf{k}}-points.

\section{Interplay between structure and magnetism}\label{StructureMagnetism} 
\begin{figure}[h]
\centering
\includegraphics[scale=0.6]{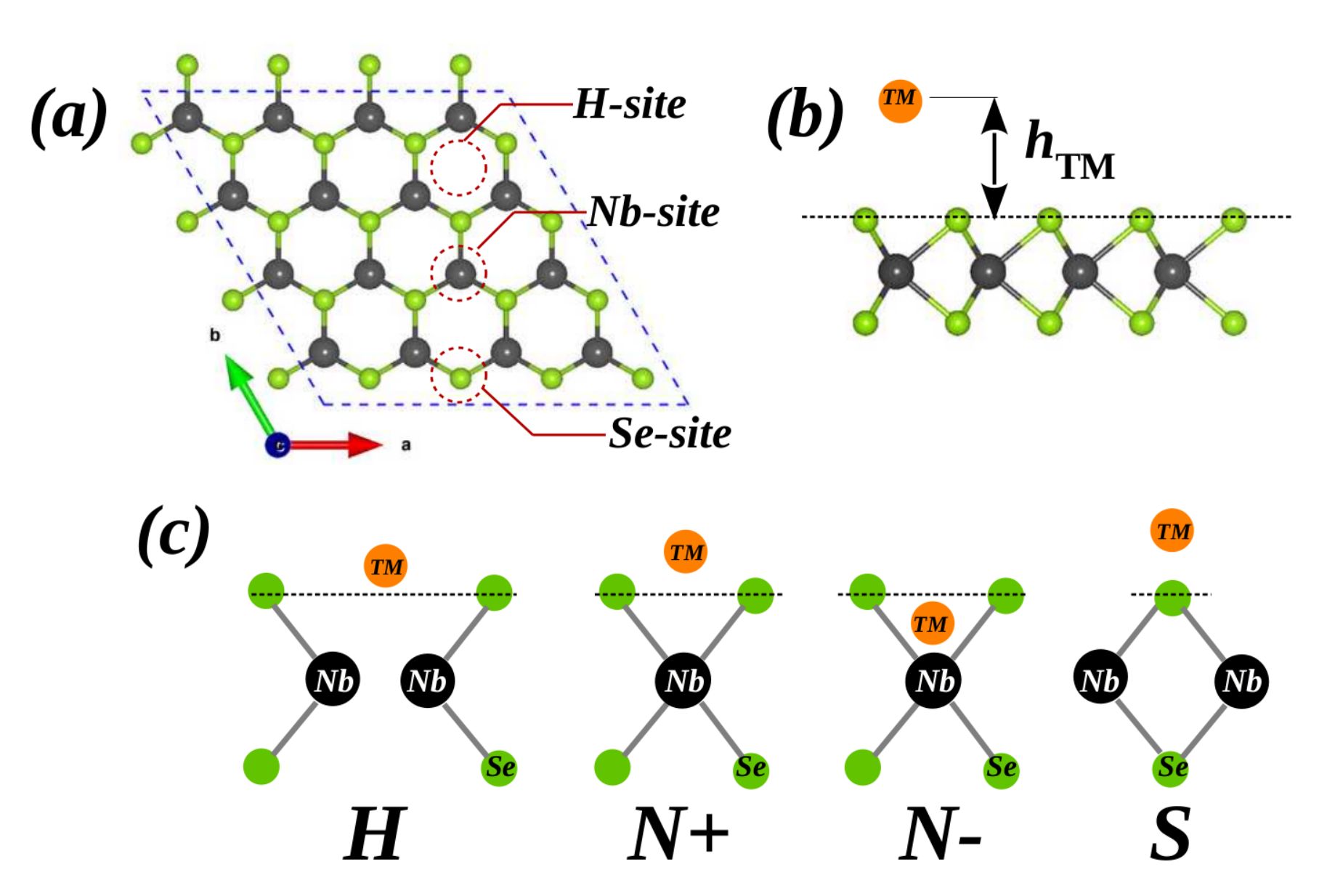}
\caption{(a) Top view of the $4 \times 4$ supercell of 1H-NbSe$_2$ monolayer, showing the 3 considered sitting positions for the TM atoms. The dotted line indicates the size of the supercell. (b) Side view of a part of the NbSe$_2$ monolayer. The topmost layer of Se atoms is shown with a dotted line, which is used as a reference for the height of the TM atoms. (c) Schematics showing the 4 possible equilibrium positions identified in DFT without spin polarization. In the N$^+$ configuration, the TM atom sits at a finite positive height above the Se layer, whereas in the N$^-$ configuration the TM atom moves closer to the Nb atoms, making the height negative.}
\label{Fig1}
\end{figure}

We performed the structural relaxation of Cr, Mn, Fe and Co adatoms on 1H-NbSe$_2$ monolayer. We focused on the $4 \times 4$ supercell, where the adatoms are sufficiently distant to investigate the long range character of the magnetic interaction. As illustrated in Fig.~\ref{Fig1}(a), we considered three possible positions for the adatoms, which are the sites on top of Nb ions or Se ions, as well as the hollow (H) site. Other possible positions, as e.g. the bridge site positions between Nb and Se atoms, can be ignored based on previous studies on similar systems~\cite{wei17prb95:075419,li17natcomm8:1958,cossu18prb98:195419,tong17advmat29:1703123,badrtdinov202dmat7:045007}. We first performed a detailed analysis of the energetic landscape without including spin polarization. These data, which are discussed more extensively in the Appendix, show that adatoms located on top of Nb can relax in two different configurations. In the first one, which we label as N$^+$, the adatom is located above the NbSe$_2$ monolayer, as expected. In the second one, which we label as N$^-$, the adatom sinks slightly below the plane of the Se ions, pushing the neighbors further away. The adatoms at the hollow site and at the Se site relax to a single configuration each, and for convenience we label them as H and S. The 4 possible configurations obtained are visualized in Fig.~\ref{Fig1}(c).

\FloatBarrier

We can now proceed to investigate how magnetism affects this scenario, by performing spin-polarized calculations in DFT. The relative energy $\Delta E$, the height of the adatom above the Se plane,  $h_{\mathsf{TM}}$, as well as the magnetic moments of the impurity and the nearest Nb atoms, respectively $\mu_{\mathsf{TM}}$ and $\mu_{\mathsf{Nb}}$, are reported in Table~\ref{Table1}. For all impurities, $h_{\mathsf{TM}}$ is increased with respect to the data without spin polarization, which is consistent with an increased bond length associated to $3d$ magnetism. This mechanism, however, does not affect the N$^-$ configuration, due to the geometrical constraints on the adatom. As a result, the N$^-$ configuration is energetically penalized, by a correction that is proportional to the magnetic moment. Hence, all the adatoms prefer to occupy the H position. 
As for the non magnetic calculations, the S configurations are always much higher in energy.
Focusing on the magnetic moments, we see an evident correlation between $\mu_{\mathsf{TM}}$  and $h_{\mathsf{TM}}$, for all adatoms. In the S configuration, the largest height corresponds to the largest moment, characterized by a small hybridization with the NbSe$_2$ monolayer. In the N$^+$ configuration, the height and the moment are slightly decreased, which indicates a larger hybridization. A further decrease is observed for the H configuration, which we already identified as the most favourable arrangement. Finally, the moment is drastically quenched in the N$^-$ configuration, due to the strong overlap with states from Nb and Se. For Fe and Co, the hybridization is so strong that the system does not even manage to form a significant local moment. A more quantitative analysis of the hybridization is provided in Appendix B. Finally, we also note that a small moment is induced on the Nb atoms. For the nearest neighbors, this moment is anti-ferromagnetically aligned for Cr and Mn, but ferromagnetically aligned for Fe and Co. This is consistent with the general behavior of these transition metals in their bulk form~\cite{kvashnin16prl116:217202,cardias17scirep7:1}. 

Next, we consider that the physics of the $3d$ electrons on adatoms is usually not well captured by standard DFT with local or semi-local functionals~\cite{wehling11prb84:235110,wei17prb95:075419,li17natcomm8:1958,cossu18prb98:195419,reed21advmatint2100463}. To remedy this problem, we performed fully relaxed DFT+U calculations, whose results are reported on the right side of Table~\ref{Table1}. 
As expected, the inclusion of an explicit Coulomb interaction term increases the localization of the TM-$3d$ electrons, which in turn results in a weaker covalent bonding with the Nb-$4d$ states. Thus, the adatom moves farther from the monolayer and $\mu_{\mathsf{TM}}$ increases, acquiring more atomic-like character. This increased moment induces a larger polarization of the Nb atoms too, i.e. a larger $\mu_{\mathsf{Nb}}$. 
Although this may seem obvious at first, one should also keep in mind that the increased localization of the TM-$3d$ states also implies a smaller hopping, and therefore a weaker exchange coupling with the Nb-$4d$ states. However, this effect does not seem to be significant in these systems. Notice also that no magnetization is seen for Co in the N$^-$ configuration, as the Coulomb interaction is not strong enough to overcome the hybridization with the neighboring orbitals, as better illustrated in the Supplementary Material (SM). 
Concerning the energetic stability, in DFT+U all adatoms prefer to arrange in the N$^+$ configuration, in place of the H configuration. This is consistent with previous studies on NbSe$_2$ and MoS$_2$ monolayers, despite the usage of supercells with different periodicities~\cite{wei17prb95:075419,cossu18prb98:195419}.
Finally, it is important to stress that our analysis did not show any evidence of metastable spin states, as e.g. those reported for adatoms on graphene or MoS$_2$~\cite{wehling11prb84:235110,wei17prb95:075419}. 
This is due to the fact that NbSe$_2$ monolayer is metallic and therefore does not favour the formation of metastable states. This is particular important for DFT+U calculations, which may lead to a plethora of local minima for atomic-like systems as adatoms~\cite{kulik15jcp142:240901}. Since the N$^+$ and N$^-$ configurations arise also without spin polarization (see Appendix A), they cannot be considered as metastable spin states.

\begin{table}[]
\caption{\label{Table1} Total energy $\Delta E$ (eV) relative to the ground state, height of the adatom $h_{\mathsf{TM}}$ ({\AA}) above the Se plane, total magnetic moment of the adatom $\mu_{\mathsf{TM}}$ ($\mu_\mathsf{B}$), and total magnetic moment of the closest Nb atom $\mu_{\mathsf{Nb}}$ ($\mu_\mathsf{B}$). The negative sign of $\mu_{\mathsf{Nb}}$ indicates that this magnetic moment is anti-parallel to the magnetic moment of the transition metal adatom. Data for both DFT and DFT+U are reported.}
\begin{indented}
\lineup
\item[]\begin{tabular}{@{}lllllllllll}
\br
&& \centre{4}{DFT} &&  \centre{4}{DFT+U} \\ \ns \ns
&& \crule{4} && \crule{4} \\
&& $\Delta E$ & $h_{\mathsf{TM}}$ & $\mu_{\mathsf{TM}}$ & $\mu_{\mathsf{Nb}}$ & & $\Delta E$ & $h_{\mathsf{TM}}$ & $\mu_{\mathsf{TM}}$ & $\mu_{\mathsf{Nb}}$ \\
\mr
\textbf{Cr} & H     & \texttt{GS} &  0.39 & 2.70 &\-0.21 & & 0.24        &  1.47 & 4.20 &\-0.28 \\
            & N$^+$ & 0.17        &  1.26 & 3.68 &\-0.20 & & \texttt{GS} &  1.41 & 4.16 &\-0.24 \\
            & N$^-$ & 0.50        &\-0.86 & 1.31 &\-0.10 & & 1.67        &\-0.84 & 2.66 &\-0.46 \\
            & S     & 1.48        &  2.21 & 4.27 &\-0.06 & & 0.89        &  2.27 & 4.43 &\-0.07 \\
\mr
\textbf{Mn} & H     & \texttt{GS} &  0.50 & 3.52 &\-0.18 & & 0.08        &  1.18 & 4.40 &\-0.26 \\
            & N$^+$ & 0.08        &  1.20 & 3.98 &\-0.15 & & \texttt{GS} &  1.30 & 4.40 &\-0.15 \\
            & N$^-$ & 0.61        &\-0.85 & 1.48 &\-0.30 & & 1.91        &\-0.78 & 3.18 &\-0.56 \\
            & S     & 1.88        &  2.29 & 4.66 &  0.02 & & 1.27        &  2.38 & 4.89 &  0.04 \\
\mr
\textbf{Fe} & H     & \texttt{GS} &  0.25 & 2.73 &  0.11 & & 0.02        &  0.49 & 3.21 &  0.16 \\
            & N$^+$ & 0.45        &  0.92 & 3.00 &  0.20 & & \texttt{GS} &  1.03 & 3.34 &  0.28 \\
            & N$^-$ & 0.18        &\-0.86 & 0.10 &  0.00 & & 1.11        &\-0.64 & 2.34 &\-0.39 \\
            & S     & 2.57        &  2.01 & 2.97 &  0.17 & & 1.84        &  2.36 & 3.79 &  0.23 \\
\mr
\textbf{Co} & H     & \texttt{GS} &\-0.17 & 1.16 &  0.15 & & 0.21        &  0.23 & 1.97 &  0.36 \\
            & N$^+$ & 0.50        &  1.01 & 1.74 &  0.11 & & \texttt{GS} &  1.12 & 1.94 &  0.20 \\
            & N$^-$ & 0.29        &\-0.77 & 0.01 &  0.00 & & 1.35        &\-0.87 & 0.02 &  0.00 \\
            & S     & 2.65        &  1.98 & 1.92 &  0.13 & & 2.30        &  2.07 & 2.22 &  0.22 \\
\mr
\end{tabular}
\end{indented}
\end{table}
\FloatBarrier

\section{Long-range inter-atomic exchange coupling}\label{ExchangeCoupling} 

\begin{figure}[h]
\centering
\includegraphics[scale=0.5]{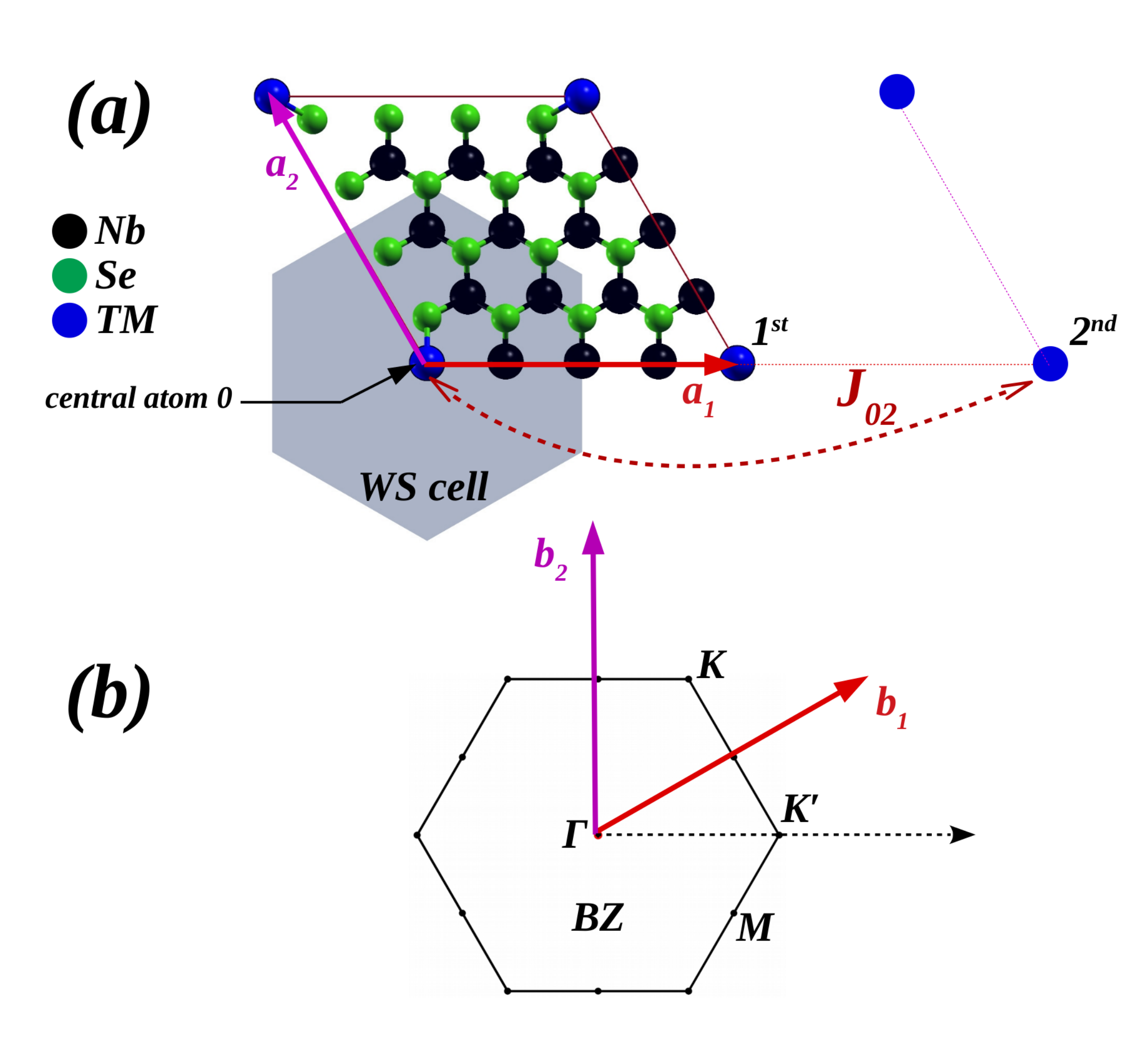}
\caption{(a) Top view of the $4\times4$ NbSe$_2$ supercell with TM adatoms in the N$^+$ configuration, including the lattice vectors $\mathbf{a}_1$ and $\mathbf{a}_2$ as well as the Wigner-Seitz (WS) cell. J$_{02}$ is the inter-atomic exchange interaction between the central TM atom and its second neighbor along the $\mathbf{a}_1$ lattice vector direction. (b) Corresponding Brillouin Zone (BZ), including the two reciprocal lattice vectors $\mathbf{b}_1$ and $\mathbf{b}_2$, as well as high symmetry points. The reciprocal space direction along $\Gamma$-K$'$ corresponds to the real space direction along $\mathbf{a}_1$.}
\label{Fig2}
\end{figure}

\begin{figure}[h]
\centering
\includegraphics[scale=0.6]{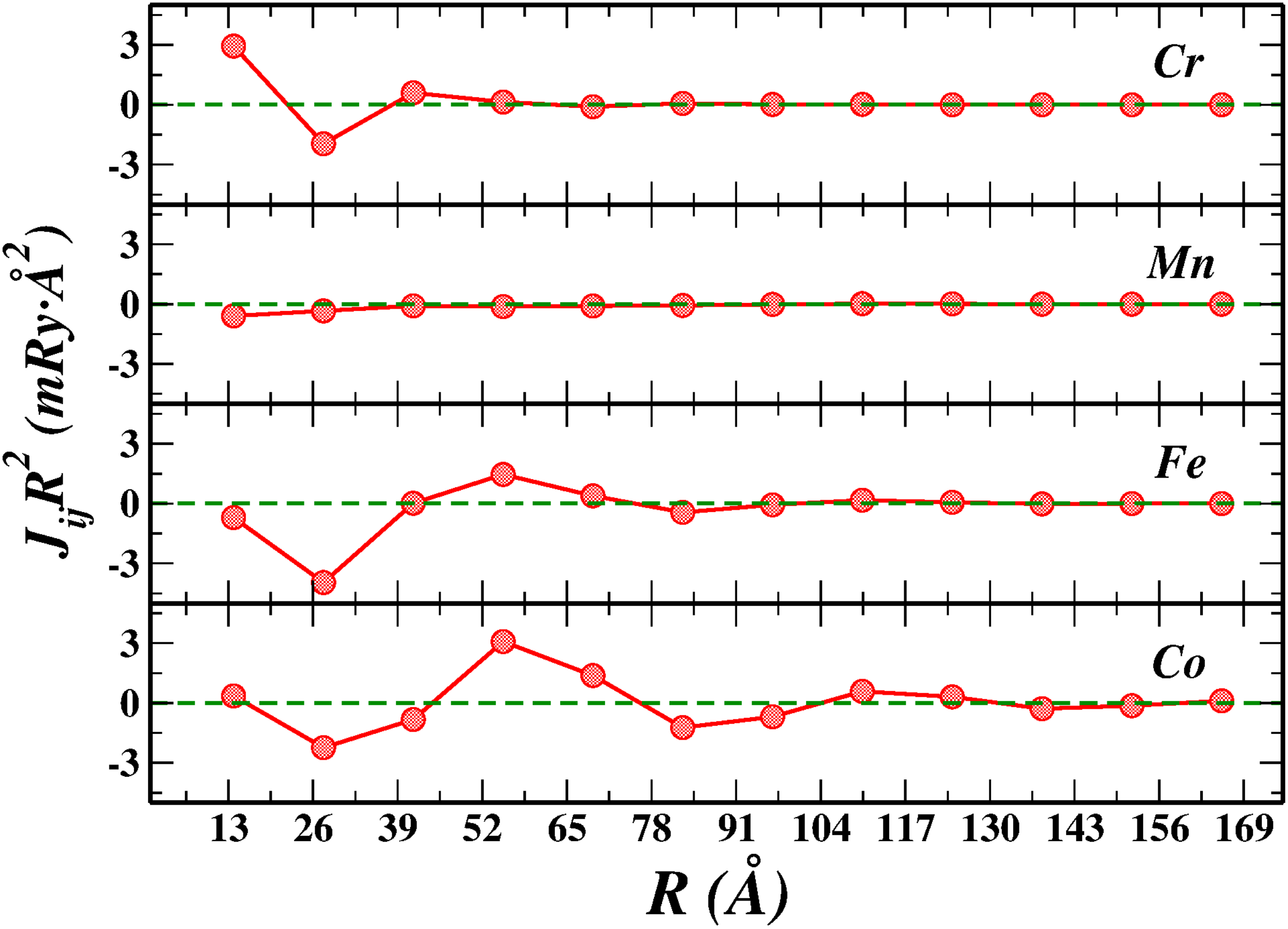}
\caption{Inter-atomic exchange interaction $J_{ij}$ between adatoms at sites $i$ and $j$ along the zigzag direction, as a function of their distance $R$. Data for Cr, Mn, Fe and Co in the N$^+$ configuration in DFT+U are shown. Note that the $J_{ij}$'s are multiplied by R$^2$ to take into account their expected asymptotic behavior and make the long range oscillations more visible.}
\label{Fig3}
\end{figure}

Having clarified the relation between structural and magnetic configurations, we proceed to investigate the long range behavior of the exchange interaction. We focus mainly on the ground state structure N$^+$, while additional data for the H configuration, which lays slightly above in energy, are shown in the SM. Since we are interested in the long range coupling, it is more convenient to discuss results obtained along the zigzag direction, $\mathbf{a}_1$ in Fig.~\ref{Fig2}(a), which ensures the maximum line density. 
Focusing on this direction, we calculated the inter-atomic exchange interactions $J_{ij}$ between adatoms at sites $i$ and $j$, for Cr, Mn, Fe and Co in the N$^+$ configuration. The results obtained from DFT+U calculations, as a function of the distance $R$ between the adatoms, are shown in Fig.~\ref{Fig3}. For a better analysis of their asymptotic scaling, the $J_{ij}$'s have been multiplied times $R^2$. This factor should account for the scaling expected for the exchange coupling between two localized moments mediated by a 2D electron gas, which is a generalization of the RKKY interaction~\cite{blundell01bookpg1567,aristov97prb55:8064}. Previous studies based on model Hamiltonians have in fact shown that the inter-atomic exchange coupling between TM adatoms on doped MoS$_2$~\cite{parhizgar13prb87:125401,mastrogiuseppe14prb90:161403} scales as $R^{-2}$ at large distances. NbSe$_2$ is metallic from the outset and in principle one may expect a similar scaling even in absence of doping.  However, the inspection of Fig.~\ref{Fig3} reveals a more complex behavior, with a decay that is much faster than a quadratic scaling. As we will see below, this is a consequence of the particular FS of these systems. The second common feature for all the adatoms reported in Fig.~\ref{Fig3} is that the inter-atomic exchange interaction oscillates between being ferromagnetic (positive sign) and anti-ferromagnetic (negative sign). This is another manifestation of the RKKY coupling and can also be connected to the topology of the FS. 
 Going more into the details of each element, we can notice that Fe and Co are characterized by oscillations of similar period as well as a similar scaling. For Cr, instead, the magnetic coupling seems to decay faster and the oscillations seem to have a slightly shorter period. Finally, the decay of the exchange interaction in Mn is so fast that we cannot really resolve the period of the oscillations for the inter-atomic distances under consideration. 
 
\FloatBarrier
\FloatBarrier

\begin{figure}[h]
\centering
\includegraphics[scale=0.3]{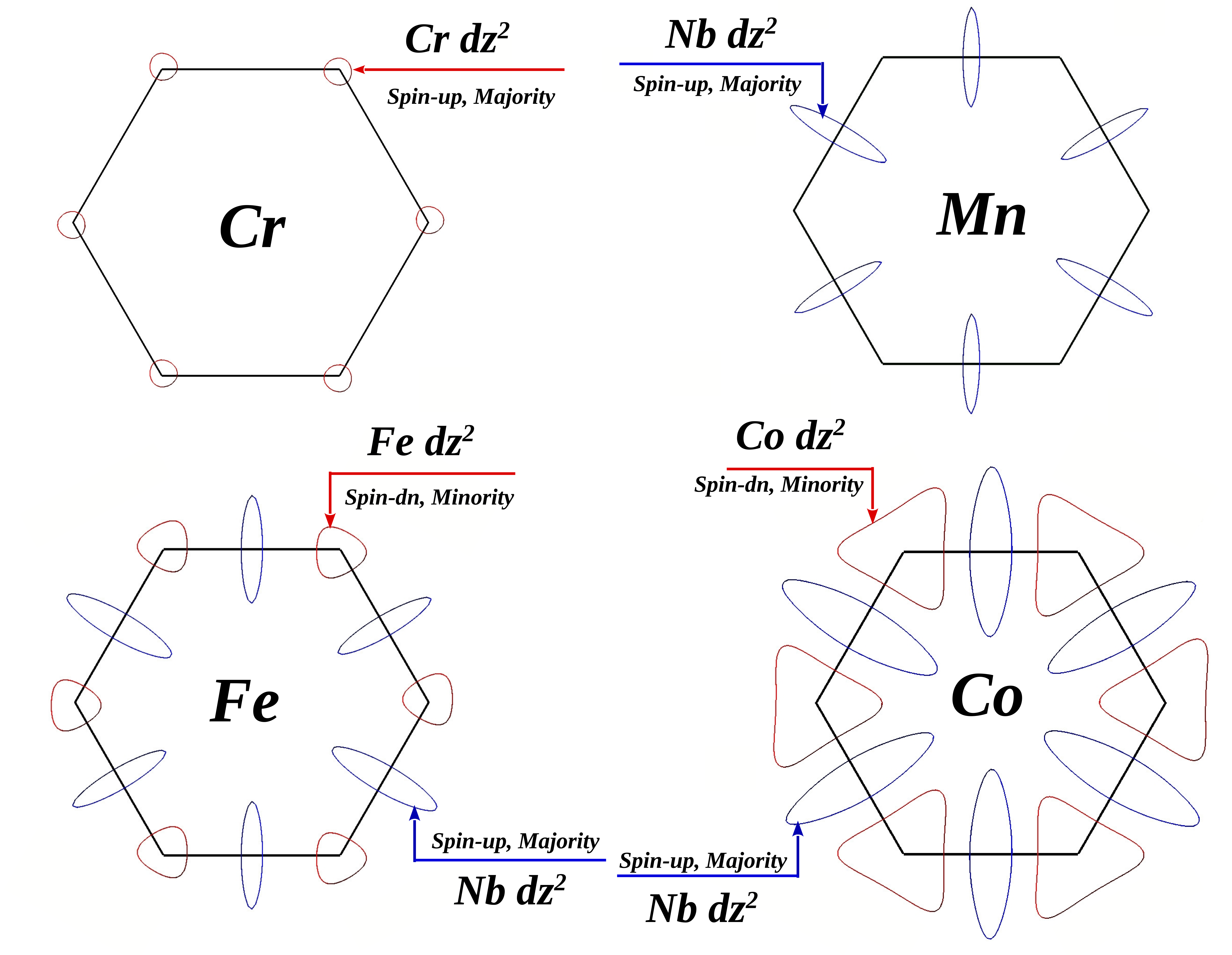}
\caption{FS of Cr, Mn Fe, and Co on monolayer NbSe$_2$ in the N$^+$ configuration, as obtained in DFT+U. Two basic features are visible: a triangular-shaped electron pocket at K and a lens-shaped hole pocket at M. The dominant spin and orbital character of the Fermi pockets is also indicated. }
\label{Fig4}
\end{figure}

\begin{figure}[h]
\centering
\includegraphics[scale=0.5]{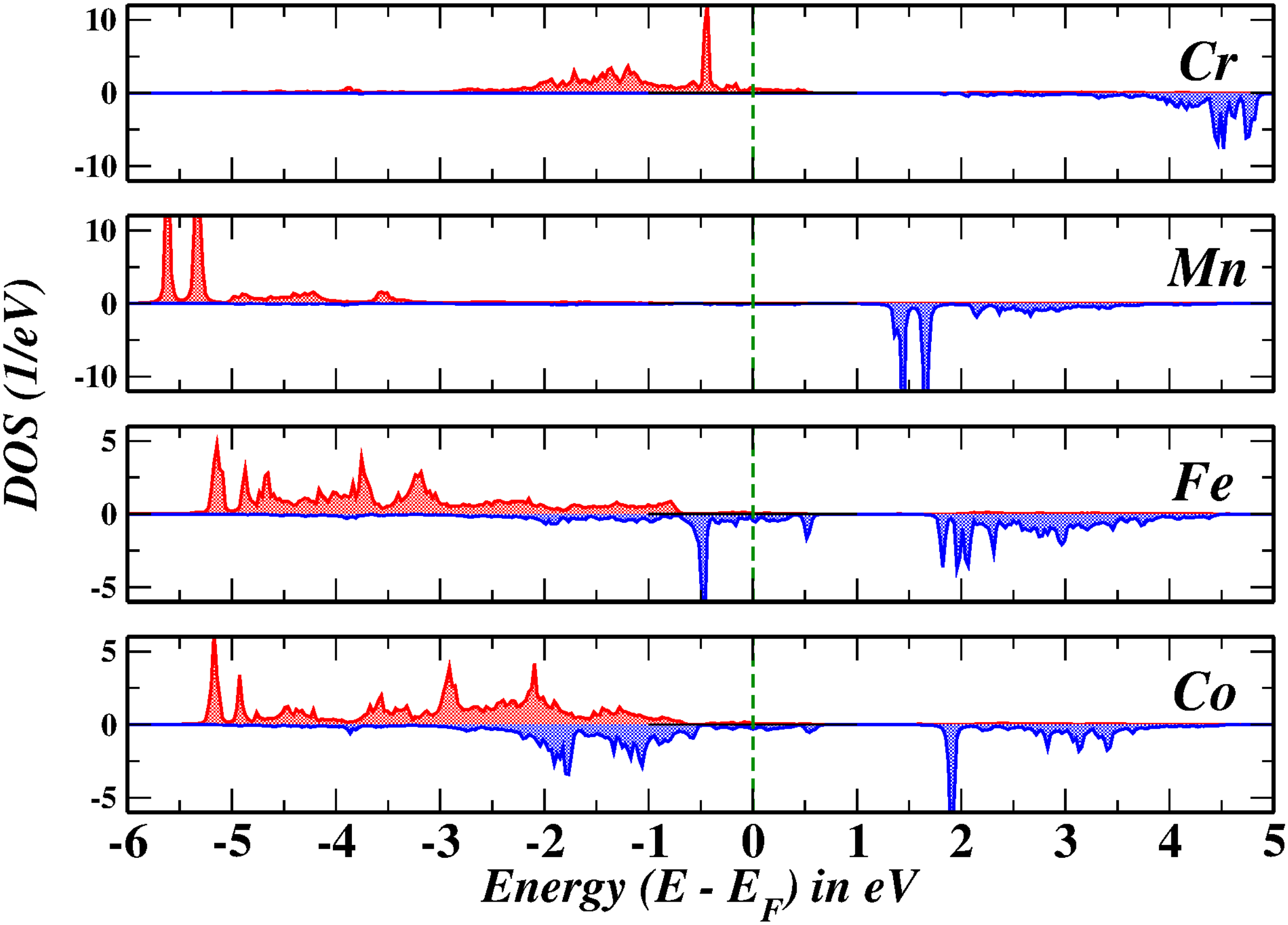}
\caption{Projected DOS for the TM-$3d$ states for selected adatoms on NbSe$_2$ as obtained in DFT+U. Positive and negative signs correspond to majority and minority spins, respectively. }
\label{Fig5}
\end{figure}

Overall, Fig.~\ref{Fig3} shows a non-trivial trend that cannot simply be interpreted in terms of a gradual filling of the $3d$ orbitals. Nevertheless, a qualitative understanding can be gained by analyzing the basic features of the density of states (DOS) and FS. The calculated FSs of all systems in their N$^+$ configuration are shown in Fig.~\ref{Fig4}. Common features include a triangular-shaped electron pocket at K or a lens-shaped hole pocket at M, or both.
The analysis of the band structure and the fat bands, included in the SM, illustrates that the electron pockets at K have major contributions from the TM-$d_{z^2}$ states whereas the hole pockets at M originate mainly from Nb-$d_{z^2}$ states. 
Besides these common features, some differences are also evident for each adatom.
The FS of the systems with Fe and Co have both electron and hole pockets. Cr has only the electron pockets at K, but no hole pockets at M, whereas the opposite happens for Mn. 
A direct comparison between the systems with Co and Fe suggests that the amplitude of the $J_{ij}$ oscillations directly depends on the volume of the Fermi pockets. In fact, Fig.~\ref{Fig4} shows that their FSs are similar, but the volume of both electron and hole pockets is smaller for Fe than for Co.
At the same time, Fig.~\ref{Fig3} shows that the $J_{ij}$s of Fe and Co have oscillations of similar period, but a smaller amplitude is observed for Fe, if compared to Co. 
In the case of Cr, we see that the FS consists of small electron pockets at K, which originate from the Cr-$d_{z^2}$ states. The hole pockets coming from Nb-$d_{z^2}$ states are missing, which indicates a weak coupling between Cr and Nb states. As a result the  $J_{ij}$ oscillations vanish very quickly for increasing inter-atomic distance. For Mn, we only have the hole pockets at the $M$ points coming from the Nb-$d_{z^2}$ states. The Mn-$d_{z^2}$ states are far from the Fermi energy, due to the large exchange splitting. This is clearly visible in the TM-$3d$ projected DOS, shown in Fig.~\ref{Fig5}. As a consequence of the limited hopping involving the TM-$3d$ states, the inter-atomic exchange interaction has a very short range.

\FloatBarrier

\FloatBarrier

\begin{figure}[t]
\centering
\includegraphics[scale=0.20]{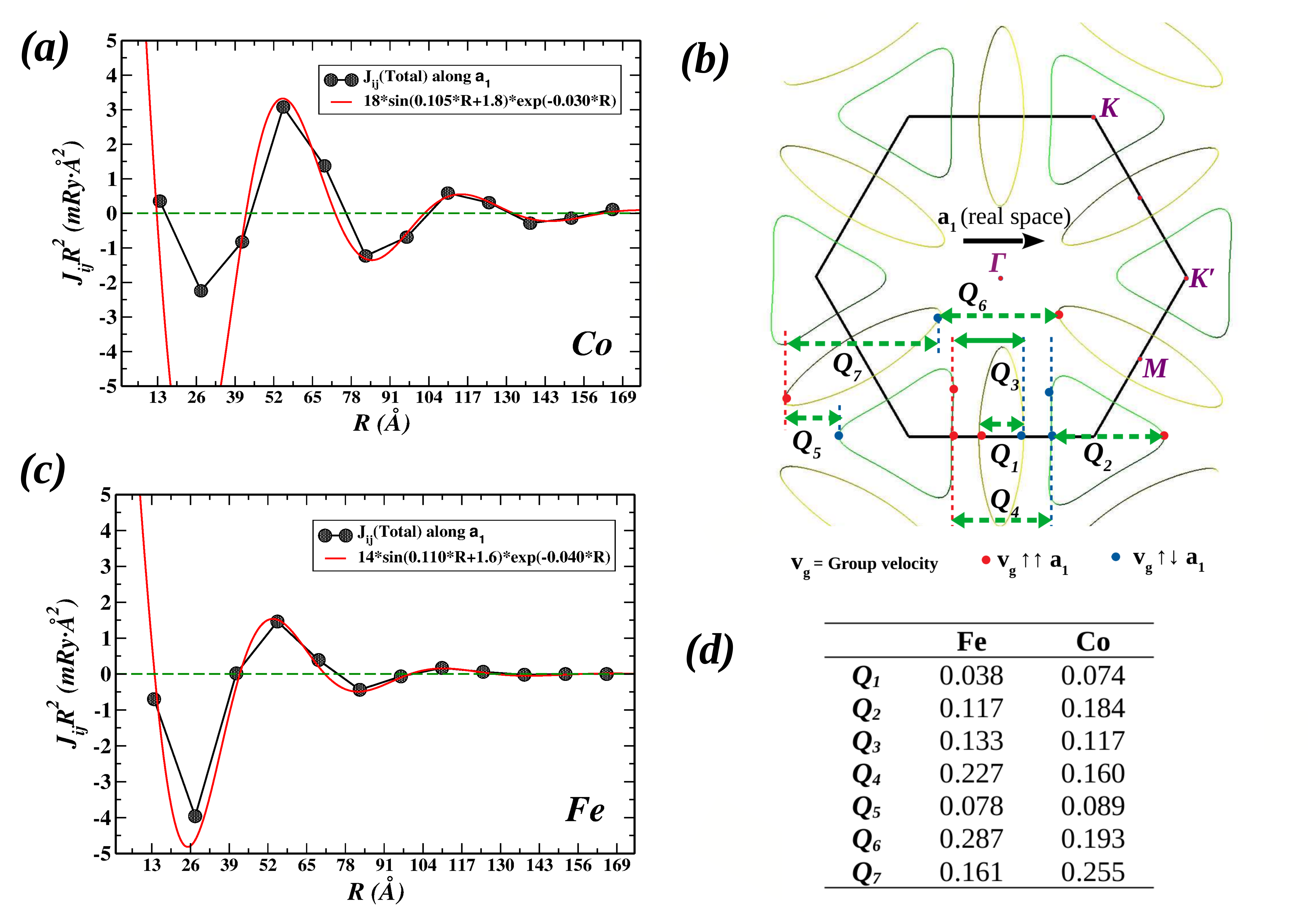}
\caption{Fitting of the inter-atomic exchange interaction $J_{ij}$ with an exponentially decaying sine wave for (a) Co and (c) Fe. (b) Possible caliper vectors identified in the FS of Co. In the case of Fe, we obtain the same vectors, albeit with a different length. (d) Table illustrating the lengths of the possible caliper vectors found for Fe and Co.}
\label{Fig6}
\end{figure}

A more comprehensive theory for understanding the behavior of the RKKY coupling between adatoms on metallic 2D materials can be formulated borrowing from the seminal works by Roth {\it{et al.}}~\cite{roth66pr149:519}, Bruno {\it{et al.}}~\cite{bruno91prl67:1602,bruno91prb43:6015,bruno92prb46:261,bruno94prb49:13231,bruno95prb52:411} and Pajda {\it{et al.}}~\cite{pajda01prb64_174402}. The concept of complex FS~\cite{bruno94prb49:13231,pajda01prb64_174402} helps us understand why the long range exchange coupling decays faster than $R^2$. As shown in Fig.~\ref{Fig4}, the contribution of the TM states to the FS include at most one spin channel, which means that a gap characterizes the other spin channel. This leads to an additional exponential decay that dominates over the quadratic scaling, as it happens e.g. in strong ferromagnets~\cite{pajda01prb64_174402,kudrnovsky04prb69:115208}. This also explains why Mn stands out as having an exchange coupling that has a much shorter range and no marked oscillations, given that its FS shows states that have no dominant TM contribution, for either spin. 
Determining the precise factor governing the decay is complicated, but insightful information can be obtained by fitting the inter-atomic exchange via an expression as $J(R) \sim A\sin{(Q*R+\phi)}R^{-2} \exp{(-\eta R)}$. Here, $A$ and $Q$ are respectively amplitude and period of the oscillations, $\phi$ is a phase factor and $\eta$ describes the exponential decay. In Fig.~\ref{Fig6}, we illustrate the results of the fitting for Fe and Co, which are the two elements for which the RKKY nature of the coupling is more evident. Although in principle one could have a superposition of more oscillatory functions~\cite{simon11prb83:224416}, the data in Fig.~\ref{Fig6} suggest that a single mode is sufficient to describe the asymptotic behavior of Fe and Co adatoms on NbSe$_2$. 
The coefficients determining the exponential decay were found to be $\eta_{\mathsf{Fe}}=0.04$ \r{A}$^{-1}$ and  $\eta_{\mathsf{Co}}=0.03$ \r{A}$^{-1}$, providing a more quantitative connection between $J_{ij}$'s and FS. 
The periods of the oscillations were instead found to be very similar, i.e. $Q_{\mathsf{Fe}}=0.110$ {\AA$^{-1}$} and $Q_{\mathsf{Co}}=0.105$ {\AA$^{-1}$}. These vectors can be traced back to the details of the FS and in particular to the possible caliper vectors~\cite{roth66pr149:519,elliot_book_pg,pajda01prb64_174402,mirbt96prb54:6382}. 
The latter have to be parallel to the $\Gamma$-K$'$ direction, i.e. parallel to the real space direction \textbf{a$_1$} that we used to calculate the $J_{ij}$'s, see Fig.~\ref{Fig2}(b). 
As illustrated in Fig.~\ref{Fig6}(b) for Co, we can identify a total of 7 possible caliper vectors that are parallel to $\Gamma$-K$'$. Their precise values for Fe and Co are reported in Fig.~\ref{Fig6}(d).  
It is evident that the caliper vectors that are in better agreement with the fitting are those labeled as $Q_3$, connecting the electron pockets at the K or K$'$ points to the hole pockets at the M points and having a value of 0.133 \r{A}$^{-1}$ and 0.117 \r{A}$^{-1}$ for respectively Fe and Co. These caliper vectors have also important nesting properties, connecting large parallel regions from the electron and hole pockets. The small discrepancy between the identified caliper vectors and those obtained from the fitting is attributed to the numerical uncertainties characterizing the latter.

\FloatBarrier

\begin{figure}[b]
\centering
\includegraphics[scale=0.15]{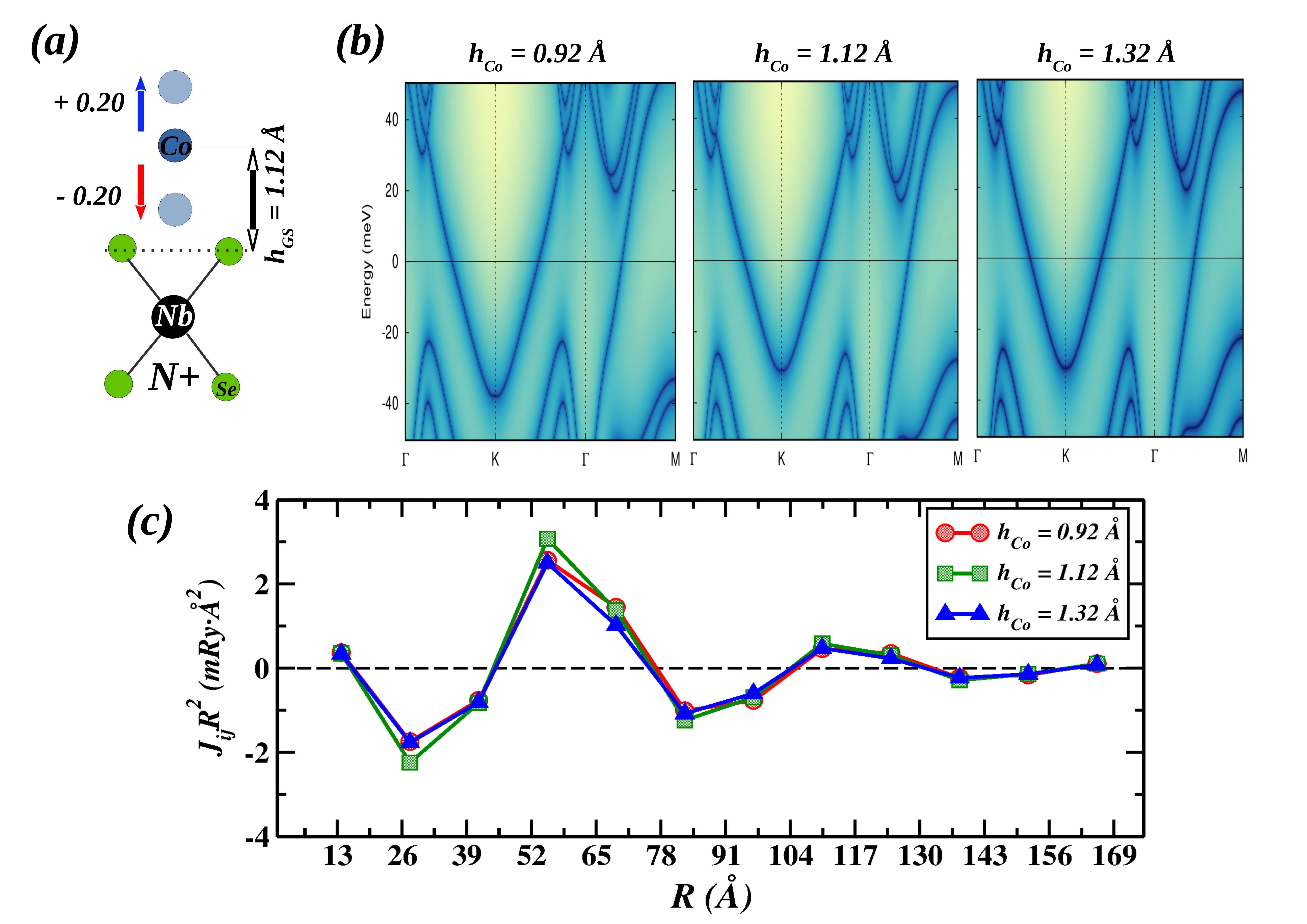}
\caption{(a) Systems where the height of the Co adatom above the Se plane is modified of $\pm 0.20$ \r{A} with respect to its equilibrium value; no relaxation of the degrees of freedom of NbSe$_2$ is performed after fixing the height. (b) Comparison of the band structure obtained for different heights of the Co atom along the important high-symmetry paths in the BZ. (c) Comparison of the inter-atomic exchange interactions $J_{ij}$'s for different heights of the Co atom.}
\label{Fig7}
\end{figure}
\FloatBarrier

When comparing the properties of the exchange coupling for Fe and Co adatoms, we should keep in mind that the difference between these two systems is not limited to the FS, but involves also the local charge and magnetic moment, which in turn affect the local hybridization. To avoid these multiple changes, we performed additional calculations for the Co system only, where the height of the impurity is varied with respect to its GS value, as shown in Fig.~\ref{Fig7}(a). Relating the changes obtained for the $J_{ij}$'s to those obtained in the FS can help us clarify if the previously identified caliper vectors are really those responsible for the long-range behavior of the RKKY coupling. 
The inter-atomic exchange interactions calculated in the systems where the Co adatom height is modified of $\pm 0.20$ {\AA} with respect to $h_{\mathsf{GS}} = 1.12$ {\AA} are reported in Fig.~\ref{Fig7}(c). It is clear that the changes induced by a different height are minimal and concern mainly the amplitude of the wave and not its period. The corresponding band structures along the relevant high symmetry directions are shown in Fig.~\ref{Fig7}(b). 
We see that increasing the height leaves the electron pocket at K substantially unvaried, while the hole pocket around M becomes smaller, due to the band moving downward. Decreasing the height, instead, leads to a larger electron pocket at K and a larger hole pocket at M, with their corresponding bands moving downward and upward, respectively. 
From the FS, we can extract the full set of caliper vectors and their variations with respect to the height, as reported in Table~\ref{Table2}. As we can see, the smallest absolute and relative variations are obtained for the vector $Q_3$, which is exactly the one we had identified above. All the other vectors experience a change of about 10$\%$, which does not correspond to what observed in Fig.~\ref{Fig7}(c). 
The reason why the caliper vector $Q_3$ remains unchanged is that it connects two parallel regions of electron and hole pockets that move toward the same direction when these pockets are both increased or decreased in volume.  
Therefore, we can conclude that this caliper vector determines the long-range behavior of the magnetic coupling for Fe and Co adatoms on NbSe$_2$.

Finally, we also performed the analysis of the inter-atomic exchange interaction of the adatoms in H configuration, corresponding to the first excited state. These data, presented in the SM, show a more complex behavior, where one cannot resolve well defined oscillatory asymptotics. This behavior can be traced back to a higher complexity of the FS, indicating the presence of two overlapping modes. The exponential decay is also present and seems to be more marked than the one observed for the N$^+$ configuration.

\begin{table}[]
\caption{\label{Table2} Possible caliper vectors $Q$ ({\AA}$^{-1}$) calculated when the height $h$ ({\AA}) of the Co adatom on NbSe$_2$ is changed. The absolute variation is calculated using the range of the values, while the relative variation is obtained from it by dividing by the GS value. No further structural relaxation is performed with respect to the GS configuration. The precise visualization of the caliper vectors is provided in Fig.~\ref{Fig6}(b).}
\begin{indented}
\lineup
\item[]\begin{tabular}{@{}llllllr}
\br
& \centre{3}{$h_{\mathsf{Co}}$} &&  \centre{2}{variation} \\ \ns \ns
& \crule{3} && \crule{2} \\
& 0.92 & 1.12 & 1.32 & & absolute & relative  \\
\mr
\textbf{$Q_1$} & 0.081 & 0.074 & 0.072 & &  0.009 & 12.2 \% \\
\textbf{$Q_2$} & 0.206 & 0.184 & 0.186 & &  0.020 & 10.9 \% \\
\textbf{$Q_3$} & 0.116 & 0.117 & 0.118 & &  0.002 &  1.7 \% \\
\textbf{$Q_4$} & 0.142 & 0.160 & 0.160 & &  0.018 & 11.3 \% \\
\textbf{$Q_5$} & 0.079 & 0.089 & 0.086 & &  0.010 & 11.2 \% \\
\textbf{$Q_6$} & 0.178 & 0.193 & 0.202 & &  0.024 & 12.4 \% \\
\textbf{$Q_7$} & 0.270 & 0.255 & 0.249 & &  0.021 &  8.2 \% \\
\mr
\end{tabular}
\end{indented}
\end{table}
\FloatBarrier

\section{Conclusions}\label{Conclusions}
In this work, we have investigated the nature of the magnetic coupling between adatoms deposited on a monolayer of NbSe$_2$. We have shown that the adatoms may occupy several stable configurations, whose energetic hierarchy depends on the method used for the computations. In DFT+U, which we consider as the method of choice among those we used, all adatoms prefer to occupy a position just on top of the Nb atoms and above the Se plane, which we named as N$^+$ configuration. The magnetic moments obtained for each adatom are found to be slightly lower than their corresponding ionic values, which reflects the presence of a finite hybridization with the metallic substrate. The calculated inter-atomic exchange couplings show that pairs of adatoms interact with each other via RKKY interaction, accompanied by a further exponential decay that is due to the absence of one of the spin channels from the FS, in the case of Cr, Fe and Co. For these elements the exchange is found to oscillate between ferromagnetic and anti-ferromagnetic character, and the period of the oscillation is found to be determined by the caliper vector connecting electron pockets at K and hole pockets at M. We have also shown that this vector is rather insensitive to changes of the height of the adatom. This implies that one can in principle manipulate the magnetic character of pair of atoms on top of NbSe$_2$ by moving them in-plane, without significant effects arising from the out-of-plane modifications. This seems to offer a better realization of the magnetic interaction control between adatoms by means of external bias voltage, recently proposed by Badrtdinov {\it{et al.}} for phosphorene~\cite{badrtdinov202dmat7:045007}.
Furthermore, considering that NbSe$_2$ is metallic and that adatoms have a single spin configuration (without low and high spins), probing their magnetic interactions via inelastic electron tunneling spectroscopy (IETS) seems much more feasible. 
Cr, Fe and Co adatoms seem easier to be measured, as they experience a stronger exchange interaction. Mn adatoms may instead involve more difficulties, having an exchange interaction that is smaller and extremely short ranged. Nevertheless, this limitation may become an advantage for controlling the spin interaction at the atomic scale and tailor more complex systems with well defined magnetic properties.

\ack
We thank Prof.~Yunkyu Bang and Prof.~Peter Wahl for useful discussions on this study.
The computational work was enabled by resources provided by the Swedish National Infrastructure for Computing (SNIC) at the National Supercomputer Centre (NSC) of Link\"oping University (Sweden) and at the High Performance Computing Centre North (HPC2N), partially funded by the Swedish Research Council through grant agreement no. 2018-05973.
Financial support from the National Research Foundation (NRF), funded by the Ministry of Science of Korea, is acknowledged by F.~C. (Grant No. 2017R1D1A1B03033465 and Grant No. 2020R1C1C1005900), as well as by S.~S. and I.~D.~M. (Grant No. 2020R1A2C101217411). I.~D.~M. is also supported by the appointment to the JRG program at the APCTP through the Science and Technology Promotion Fund and Lottery Fund of the Korean Government, as well as by the Korean Local Governments, Gyeongsangbuk-do Province and Pohang City. Y.~O.~K. acknowledges the financial support from the Swedish Research Council VR (project No. 2019-03569) and G\"oran Gustafsson Foundation.

\appendix
\section*{Appendix A}
Here we focus on the energetic landscape obtained in DFT without considering any spin polarization. In Table~\ref{TableA1}, we illustrate the energy and equilibrium height of all the investigated adatoms positioned on the NbSe$_2$ monolayer. The positions are labeled as explained in the main text and illustrated in Fig.~\ref{Fig1}(c). We first notice that the energy of the S configuration is much higher than those of the other configurations, for all adatoms, and is also accompanied by the largest height. 
This suggest that a large energy gain results from the strong covalent interaction between the TM and Nb atoms, which is present for all configurations except S. 
As a matter of fact, one can see a strong correlation between the equilibrium height and the relative energy with respect to the ground state. 
As discussed in the main text, there are two possible equilibrium heights for adatoms located on top of the Nb site. These two stable configurations arise from the competition between two possible microscopic interactions.
One of them is the aforementioned covalent interaction between the TM atom and the Nb atoms. The other one is due to the steric effects between the TM atom and the topmost Se layer.
The competition between these two types of interaction results in a potential well with two minima with respect to the variation of the height of the TM atom along the $z$-direction, perpendicularly to the monolayer. The location of the two minima is on either side of the top Se layer, separated by the potential barrier resulting from the steric effects. Therefore, we obtain two different geometries at the Nb site, which were labeled as N$^+$ and N$^-$ in the main text. 
As expected from the large difference in height, the  N$^-$ configuration is more favourable in energy than the N$^+$ configuration, when spin polarization is not considered.
Table~\ref{TableA1} shows that the H configuration and the N$^-$ configuration are very close in both height and energy. In fact, Cr and Co prefer the H structure, while Mn and Fe prefer the N$^-$ structure.

\begin{table}[]
\caption{\label{TableA1} Total energy $\Delta E$ (eV) relative to the ground state and height of the adatom $h_{\mathsf{TM}}$ ({\AA}) above the S plane, as obtained in DFT without spin polarization.}
\begin{indented}
\lineup
\item[]\begin{tabular}{@{}llllllllll}
\br
& \centre{4}{$\Delta E$} &&  \centre{4}{$h_{\mathsf{TM}}$} \\ \ns \ns
& \crule{4} && \crule{4} \\
& H & N$^{+}$ & N$^{-}$ & S & & H & N$^{+}$ & N$^{-}$ & S  \\
\mr
\textbf{Cr}   &     \texttt{GS} &  1.26     &  0.17     &  4.78  & &\-0.03   &  1.07   &\-0.87   &  1.87  \\
\textbf{Mn}   &         0.06    &  1.24     &\texttt{GS}&  4.68  & &\-0.22   &  1.00   &\-0.88   &  1.85  \\
\textbf{Fe}   &         0.08    &  1.22     &\texttt{GS}&  4.26  & &\-0.28   &  0.93   &\-0.86   &  1.87  \\
\textbf{Co}   &     \texttt{GS} &  0.85     &  0.28     &  3.40  & &\-0.25   &  0.89   &\-0.76   &  1.91  \\
\mr
\end{tabular}
\end{indented}
\end{table}
\FloatBarrier

\section*{Appendix B}
\begin{figure}[h]
\centering
\includegraphics[scale=0.3]{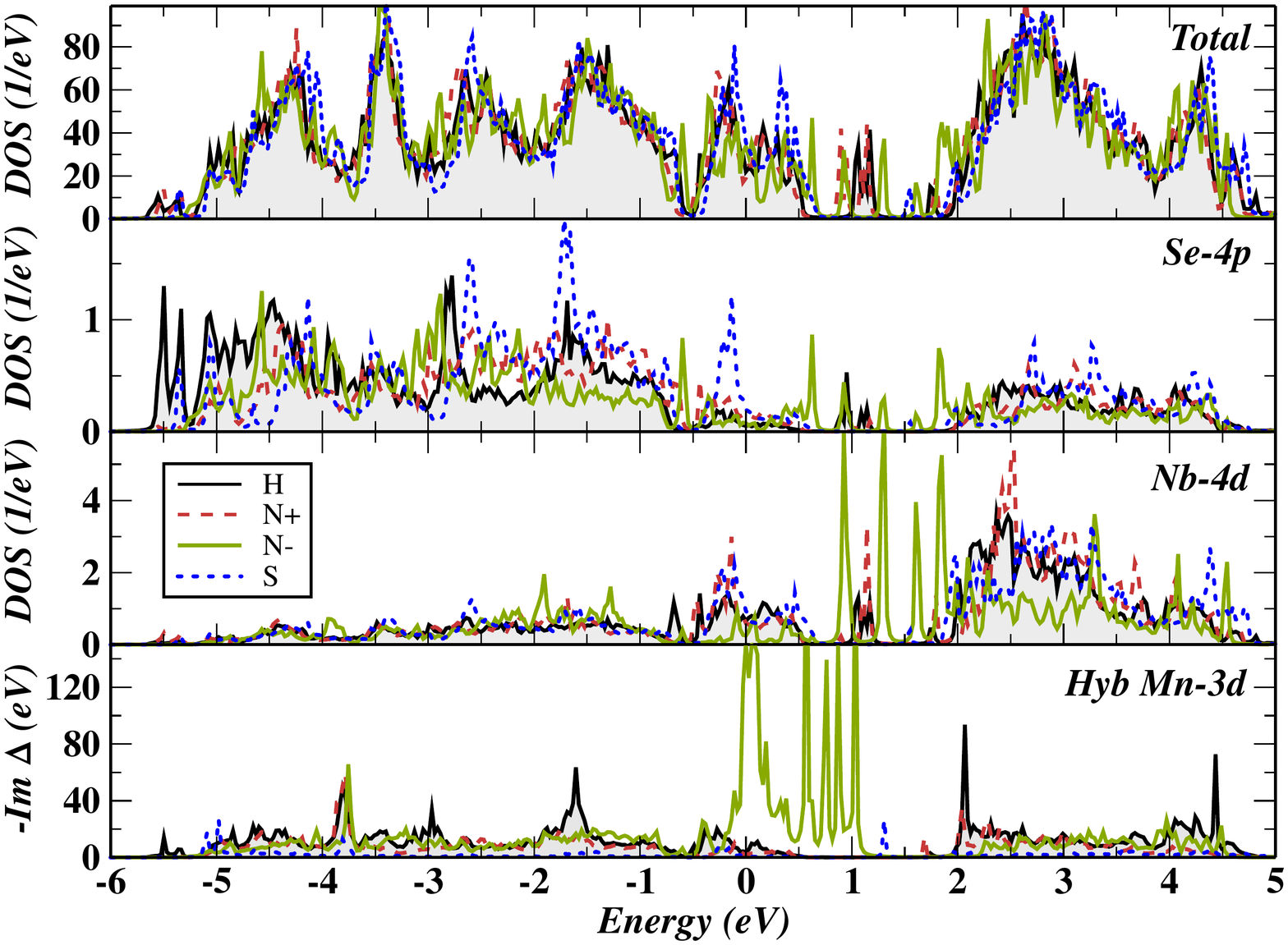}
\caption{Total density of states, projected density of states of the $4p$ states of the nearest Se atoms and the $4d$ states of the nearest Nb atoms, as well as hybridization function of the Mn-$3d$ states for the 4 considered sites of the impurity. Data obtained from spin-polarized DFT calculations.}
\label{Fig8}
\end{figure}

In this section we provide a more quantitative analysis of the hybridization of the TM-$3d$ states with their chemical environment. To this aim, we will employ the local hybridization function $\Delta(E)$, which can be obtained directly from the local Green's function, projected on a given set of local orbitals~\cite{kotliar06rmp78:865}. Recent works have shown that the hybridization function can provide significant insight into the physical properties of various compounds~\cite{herper17prm1:033802,tomczak20prb101:035116,herper20jpcm32:215502}. For sake of simplicity we will limit our analysis to Mn impurities, as the other impurities show a similar behavior. 
The trace of the imaginary part of $\Delta(E)$ for the Mn-$3d$ states, as obtained in spin-polarized DFT, is illustrated in the bottom panel of Fig.~\ref{Fig8}. The 4 different curves correspond to the 4 possible configurations discussed in the text. The other three panels of Fig.~\ref{Fig8} show the total density of states, as well as the projected densities of the $4p$ states of the nearest Se atoms and the $4d$ states of the nearest Nb atoms. Focusing on $\Delta(E)$, we observe that the smallest hybridization is found for the S configuration (blue dotted line), which is consistent with having the largest distances from Nb and Se atoms. A slightly larger hybridization is observed for the N$^+$ configuration (red dashed line), mainly determined by the Se-$4p$ states. A more structured hybridization characterizes the H configuration (black line), which is also the ground state in spin-polarized DFT. The largest peak is found at -1.5 eV and can find correspondence in the Se-$4p$ states in the same energy range. Finally, the largest hybridization is found for the N$^-$ configuration (green line) and the most important contribution is visible around the Fermi energy. In this energy range, there is a clear correspondence with Nb-$4d$ states, Se-$4p$ states, as well as other states arising from different electronic shells and visible in the total density of states. The strength of the hybridization is also found to be proportional to the difference observed between the calculated Mn magnetic moment, reported in Table~\ref{Table1}, and its expected ionic value. The trends described above are also observed in DFT+U. although the stronger localization of the TM-$3d$ states leads to a smaller hybridization function overall.

\FloatBarrier

\section*{References}
\bibliographystyle{iopart-num}
\bibliography{strings,kylie}

\end{document}



 \title[]{Supplementary material for the manuscript entitled\\``Magnetism between magnetic adatoms on monolayer NbSe$_2$''}
\author{S. Sarkar} 
\author{F. Cossu}
\author{P. Kumari}
\author{A.~G. Moghaddam}
\author{A. Akbari}
\author{Y.~O. Kvashnin}
\author{I. {Di Marco}}

\date{\today}

\begin{abstract}
This supplementary material contains band structures and fat bands analyses for the Co, Mn, Fe and Co adatoms on monolayer NbSe$_2$ in the N$^+$ configuration. Furthermore, for Co, the Fermi surfaces obtained by varying the height of the adatoms are included. Finally, the band structures, Fermi surfaces and inter-atomic exchange interaction for all the adatoms in the H configuration are also shown.
\end{abstract}
\maketitle

\begin{figure}[ht]
  \includegraphics[scale=0.45]{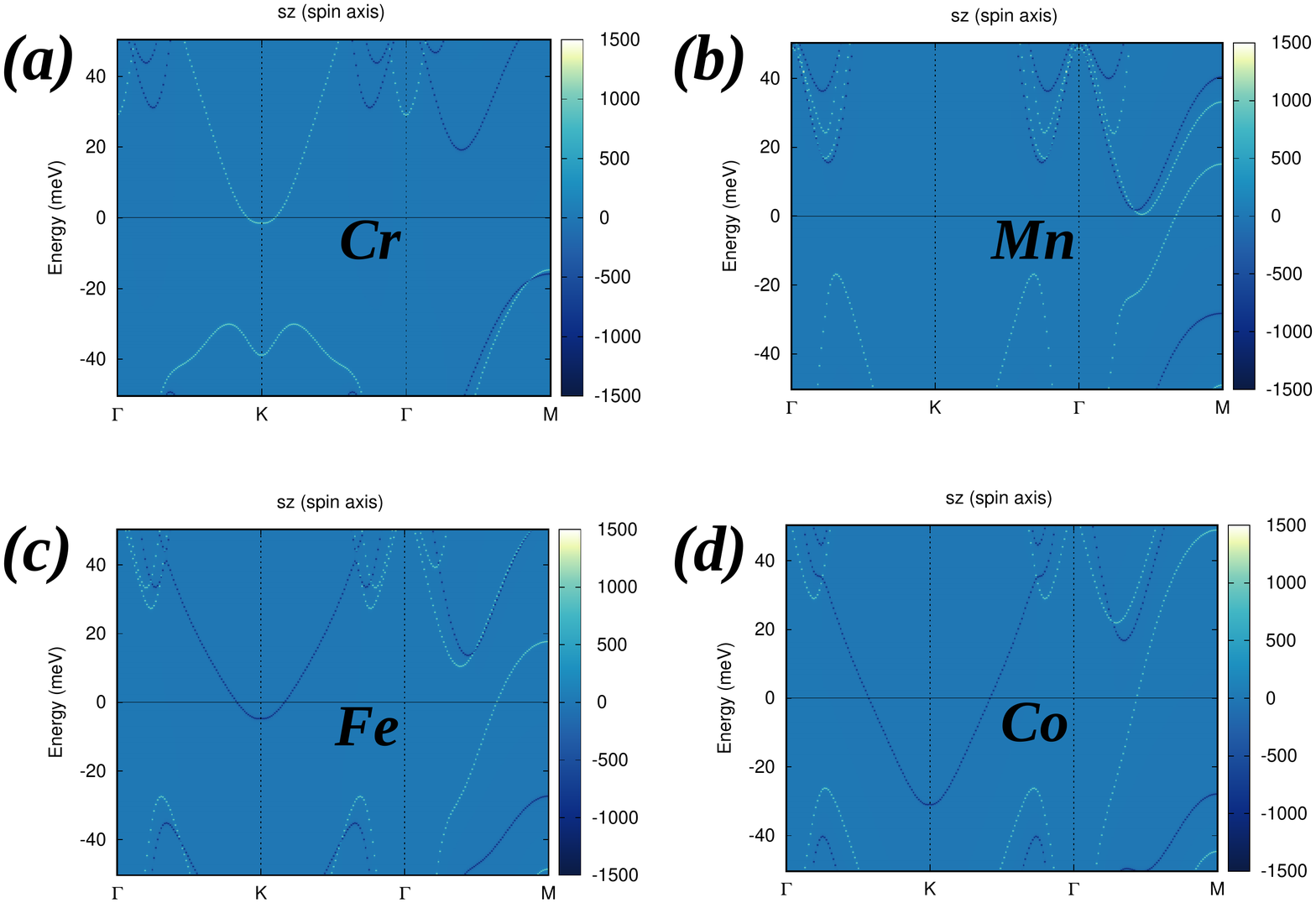}
  \label{FigS1}  
  \caption{(a) - (d) Spin polarized (color map) band structure of Cr, Mn, Fe and Co adatoms on monolayer NbSe$_2$ in the N$^+$ configuration, as obtained from DFT+U calculations with RSPt.}
\end{figure}

\begin{figure}[ht]
  \includegraphics[scale=0.25]{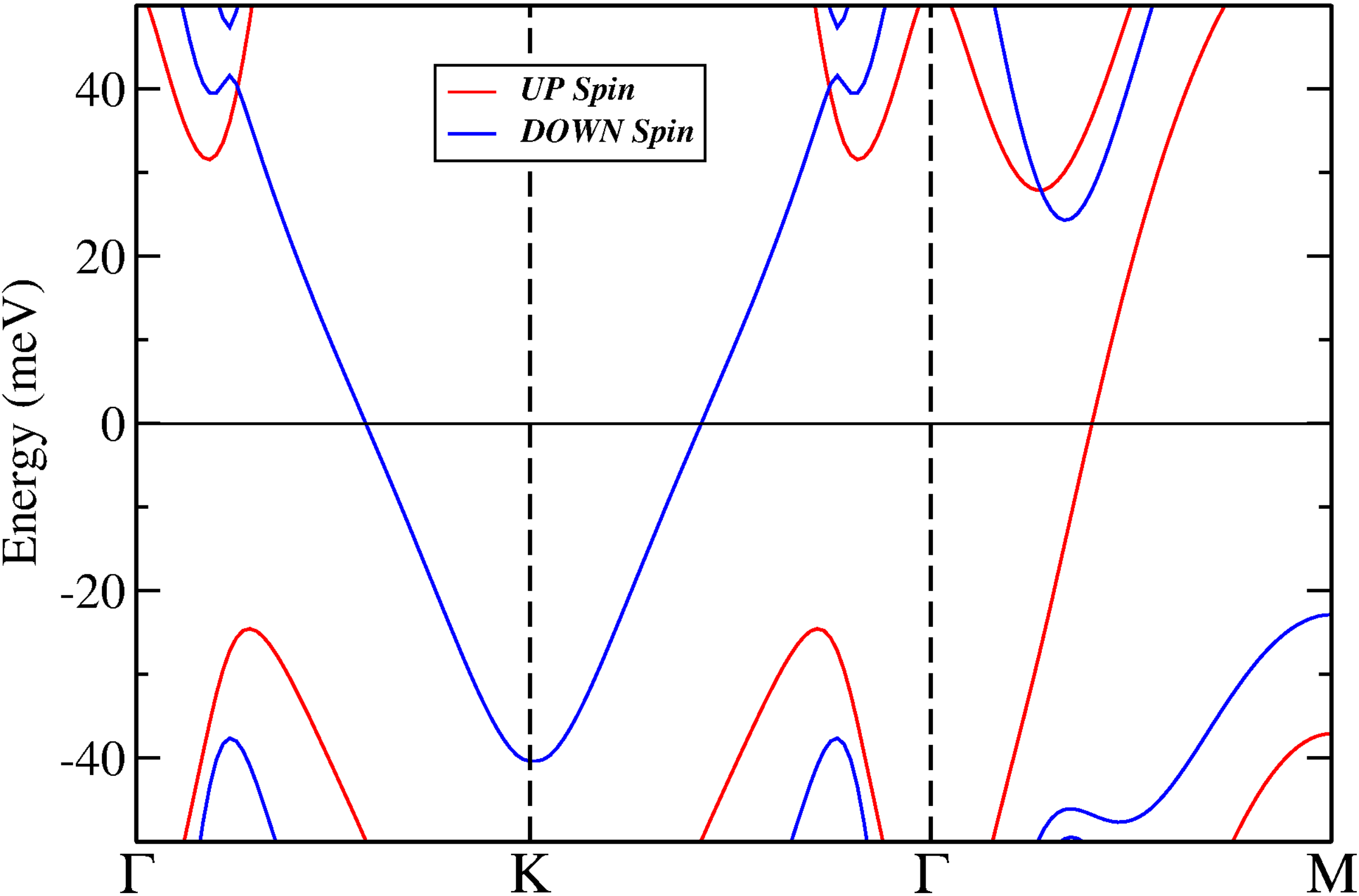}
  \label{FigS2}  
  \caption{Spin polarized band structure of Co on monolayer NbSe$_2$ in the N$^+$  configuration, as obtained from DFT+U calculations with VASP. The agreement with the RSPt data is very good.}
\end{figure}

\begin{figure}[ht]
  \includegraphics[scale=0.4]{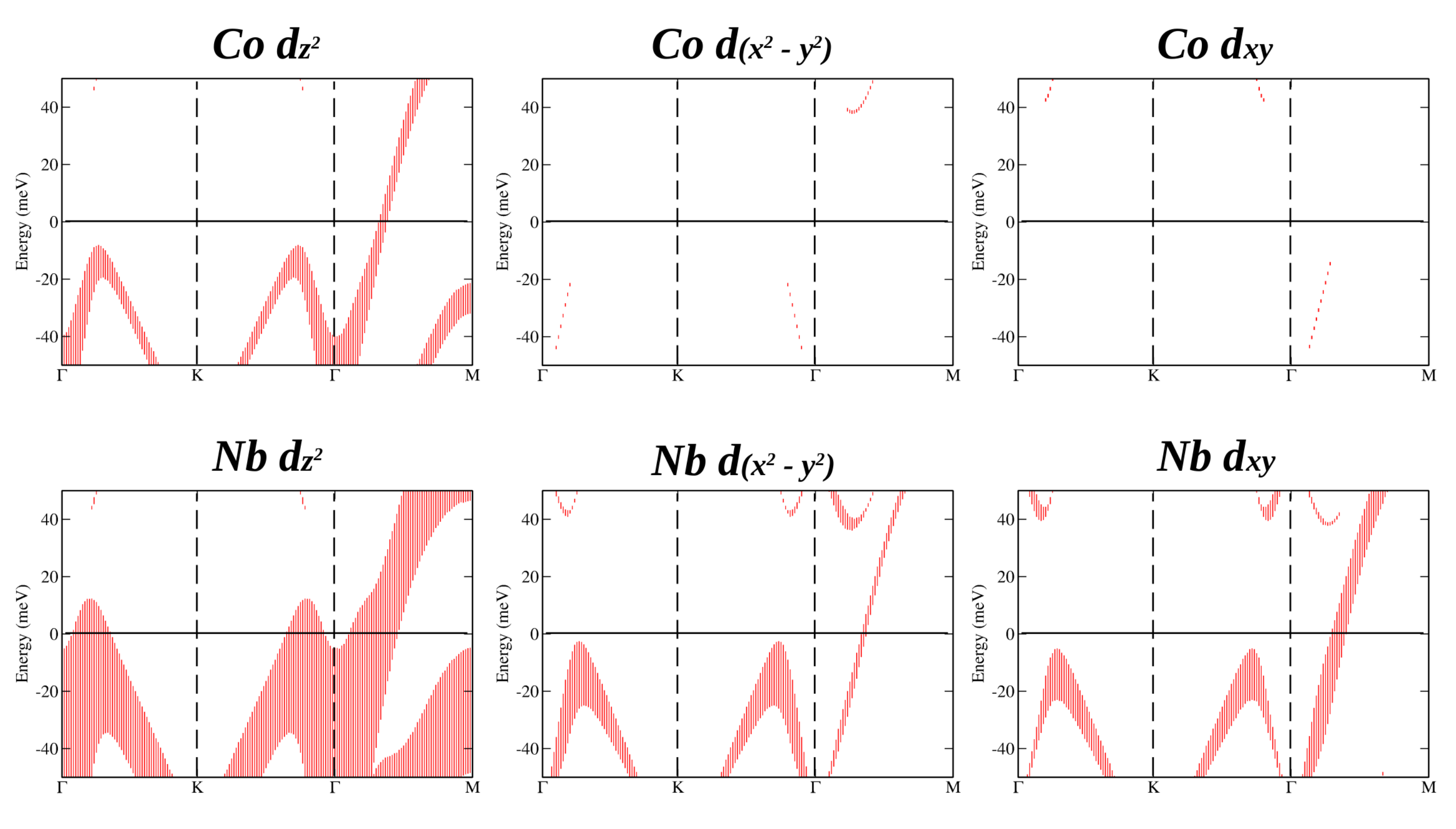}
  \label{FigS3}  
  \caption{Fatband analysis of the spin-up Co-$3d$ and Nb-$4d$ states for Co adatoms on monolayer NbSe$_2$ in the N$^+$ configuration to check the orbital character of the hole pocket at M. We see that the major contribution comes from the Nb-$d_{z^{2}}$ states with some contributions also coming from Co-$d_{z^{2}}$ and Nb-$d_{x^{2}}$ states due to hybridization. In this energy region, there are no states with $d_{xz}$ and $d_{yz}$ symmetry, for both Co and Nb. Results from DFT+U calculations with VASP.}
\end{figure}

\begin{figure}[ht]
  \includegraphics[scale=0.4]{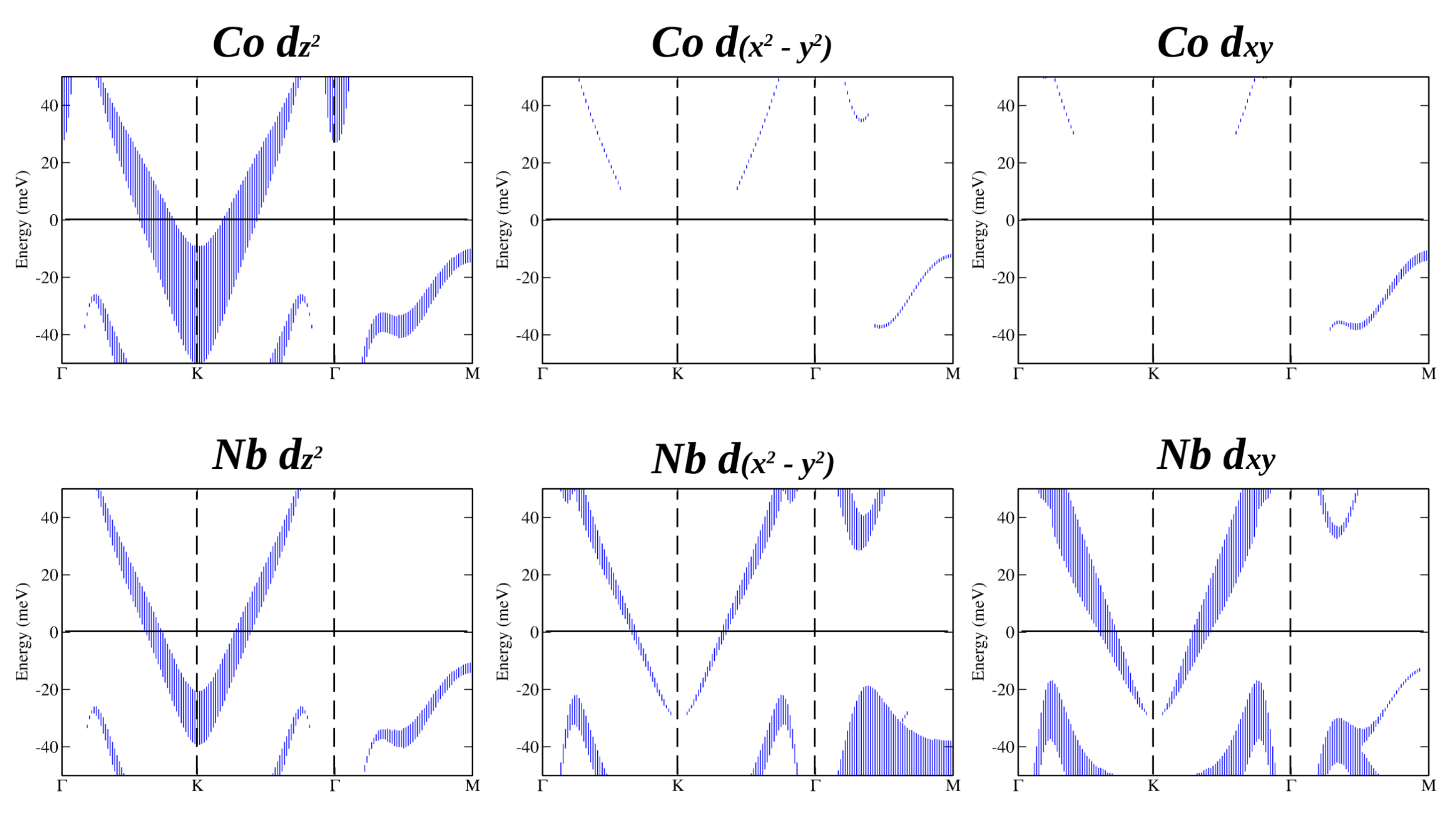}
  \label{FigS4}  
  \caption{Fatband analysis of the spin-down Co-$3d$ and Nb-$4d$ states for Co adatoms on monolayer NbSe$_2$ in the N$^+$ configuration to check the orbital character of the hole pocket at K. We see that the major contribution comes from the Nb-$d_{z^{2}}$ states with some contributions also coming from Co-$d_{z^{2}}$ and Nb-$d_{x^{2}}$ states due to hybridization. In this energy region, there are no states with $d_{xz}$ and $d_{yz}$ symmetry, for both Co and Nb. Results from DFT+U calculations with VASP.}
\end{figure}

%
%
%

\begin{figure}[ht]

    \includegraphics[scale=0.3]{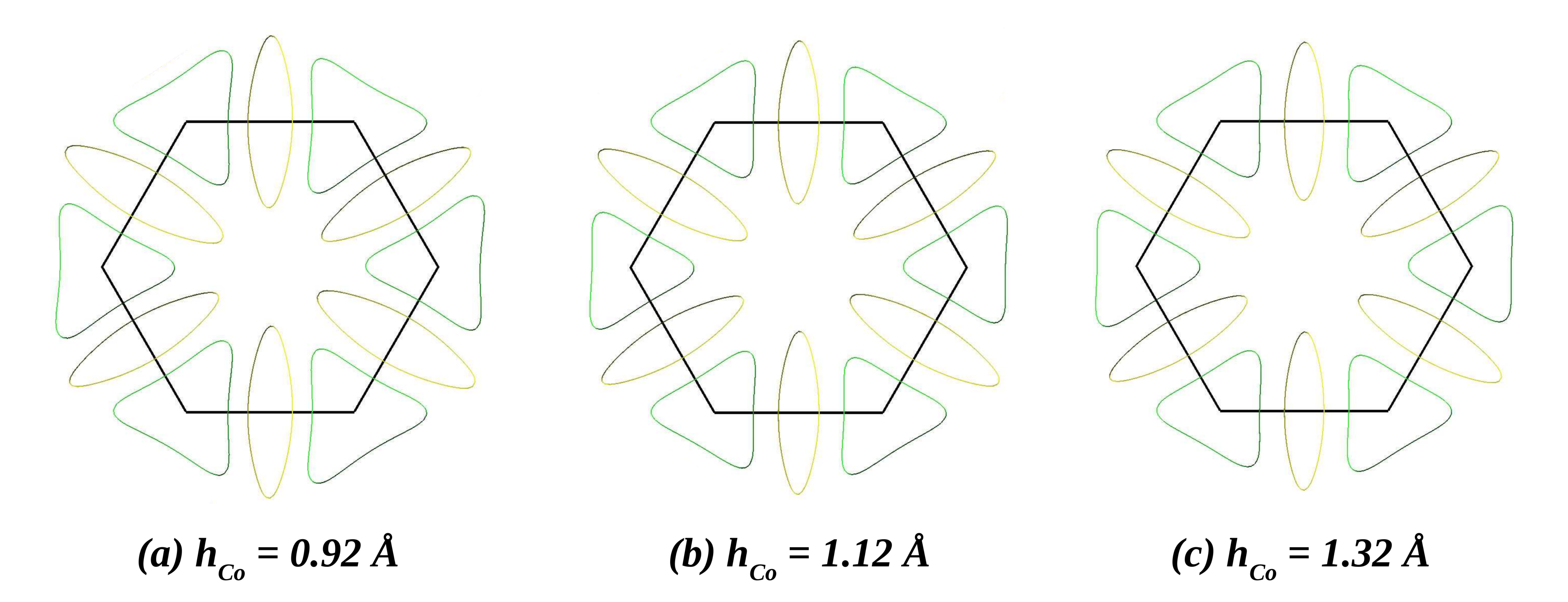}

      \label{FigS6}  
  \caption{Fermi surface of a Co adatom on monolayer NbSe$_2$ in the N$^+$ configuration, as obtained in DFT+U with RSPt. Different heights are considered, with the ground state corresponding to panel (b). See main text for details.}
\end{figure}

\begin{figure}[ht]
  \includegraphics[scale=0.4]{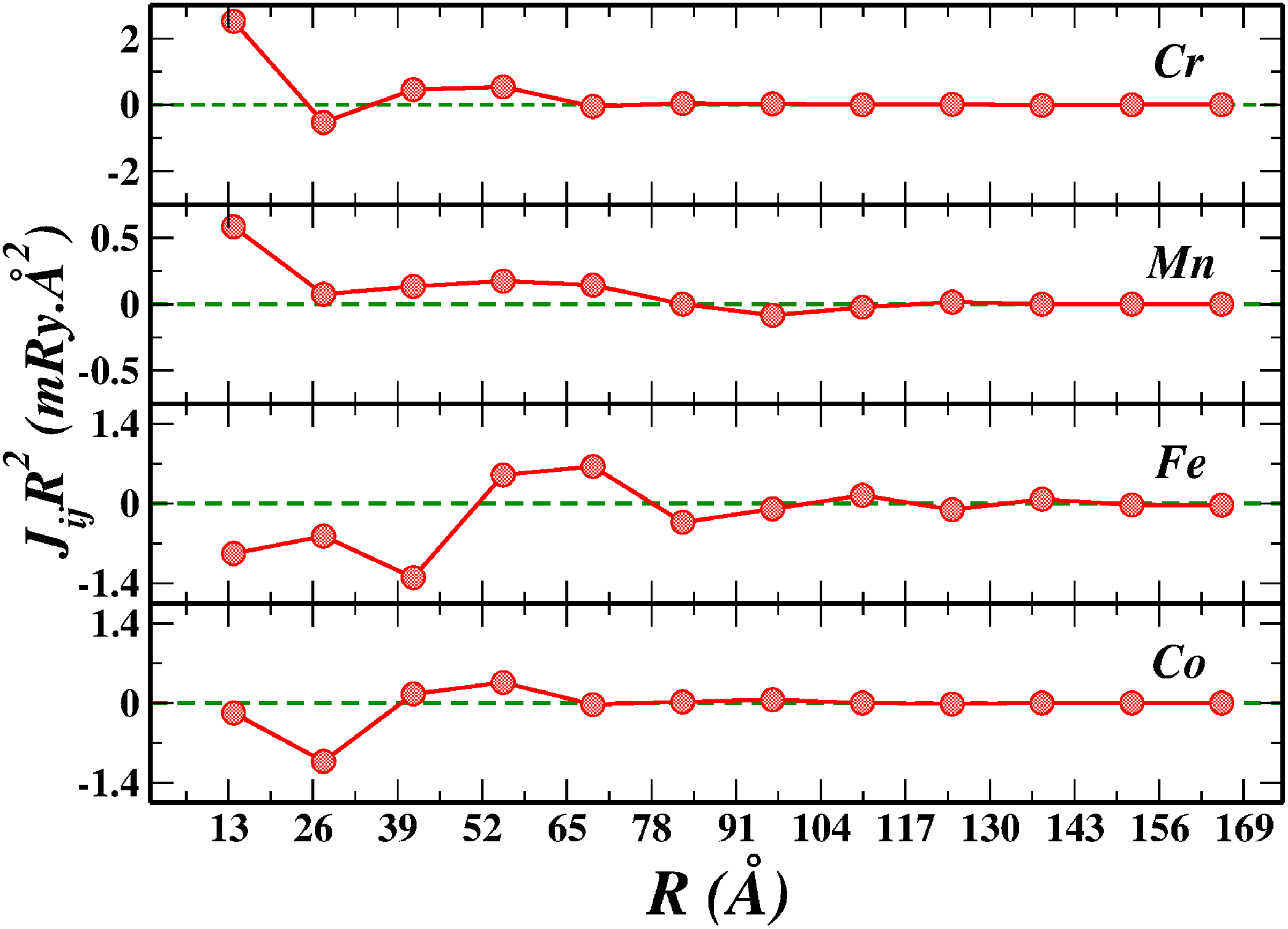}
  \label{FigS6}  
  \caption{Inter-atomic exchange interaction $J_{ij}$ between adatoms at sites $i$ and $j$ along the zigzag direction, as a function of their distance $R$. Data for Cr, Mn, Fe and Co in the H configuration in DFT+U are shown. Note that the $J_{ij}$'s are multiplied by R$^2$ to take into account their expected asymptotic behavior and make the long range oscillations more visible.}
\end{figure}

\begin{figure}[ht]
  \includegraphics[scale=0.4]{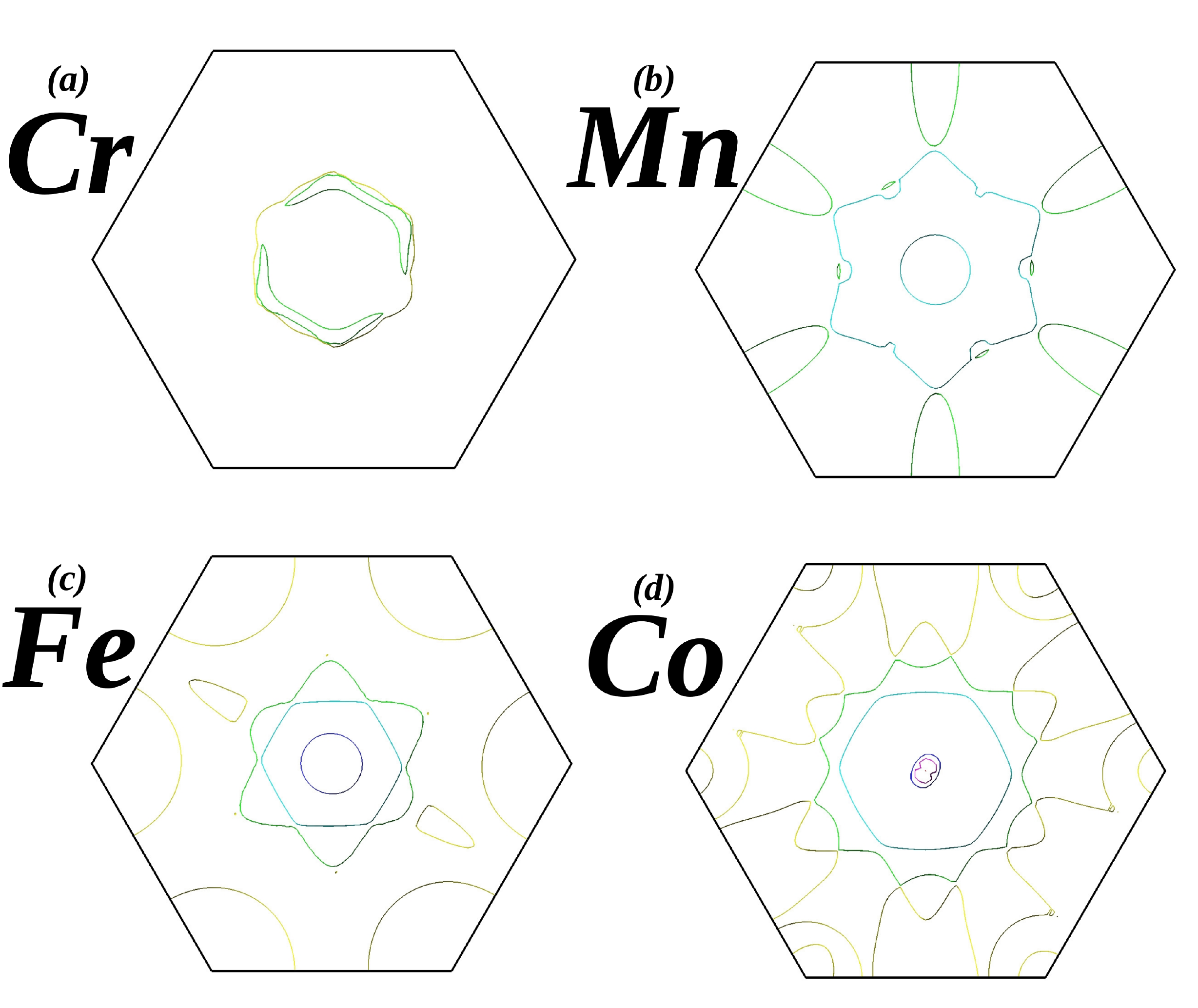}
  \label{FigS7}  
  \caption{Fermi surfaces of Cr, Mn Fe, and Co adatoms on monolayer NbSe$_2$ in the H configuration, as obtained in DFT+U with RSPt.}
\end{figure}

\begin{figure}[ht]
  \includegraphics[scale=0.45]{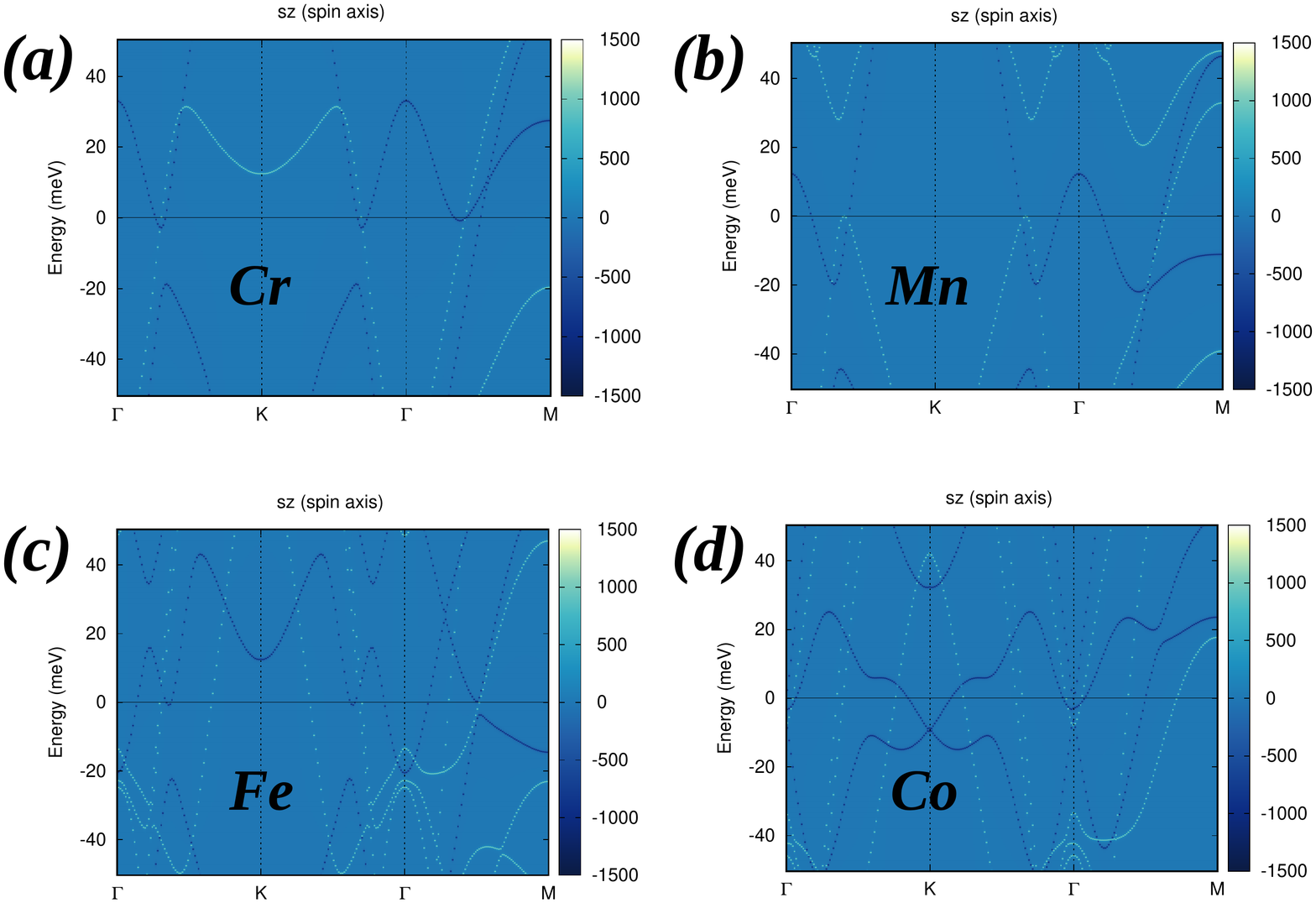}
  \label{FigS8}  
  \caption{(a) - (d) Spin polarized (color map) band structure of Cr, Mn, Fe and Co adatoms on monolayer NbSe$_2$ in the H configuration, as obtained from DFT+U calculations with RSPt. }
\end{figure}